\documentclass[letterpaper,twocolumn,10pt,anonymous]{article}
\usepackage[colorlinks=true,linkcolor=blue,breaklinks=True,citecolor=brown,urlcolor=blue]{hyperref}
\usepackage{usenix,epsfig,endnotes}
\usepackage{lipsum}
\usepackage{soul}
\usepackage{amssymb}
\usepackage{multirow}
\usepackage{makecell}
\usepackage{booktabs}
\usepackage{colortbl}
\usepackage{threeparttable}
\usepackage{array}
\usepackage{subfig}
\usepackage{svg}
\usepackage{tikz}

\usepackage[most]{tcolorbox}[breakable]
\newtcolorbox{takeaway}{colframe=black,colback=gray!15,boxrule=1pt,arc=2pt,left=2pt,right=2pt,top=1pt,bottom=1pt,before skip=1em, after skip=0em }

\newtcolorbox{cooltextbox}[1][]{%
    colback=black!5,
    colframe=black!5,
    notitle,
    sharp corners,
    borderline west={0pt}{0pt}{red!80!black},
    enhanced,
    breakable,
    left=0pt,
    right=0pt,
    top=0pt,
    bottom=0pt
    }
    
\newtcolorbox{position}[1][]{%
    colback=black!5,
    colframe=black!5,
    notitle,
    sharp corners,
    borderline west={0pt}{0pt}{red!80!black},
    enhanced,
    breakable,
    left=0pt,
    right=0pt,
    top=0pt,
    bottom=0pt
    }

\newcommand{\paragraphb}[1]{\noindent{\bf #1} }
\newcommand{\paragraphbe}[1]{\vspace{0.03in} \noindent{\bf \em #1} }
\newcommand{\jiahao}[1]{\textsf{\color{cyan}{[{Jiahao: #1}]}}}

\begin{document}

%don't want date printed
\date{}

%make title bold and 14 pt font (Latex default is non-bold, 16 pt)
\title{
Unveiling the Security Risks of Federated Learning \\ in the Wild: From Research to Practice
}

\author{
{\rm Jiahao Chen$^{1,\dagger}$, Zhiming Zhao$^{1,\dagger}$, Yuwen Pu$^{2}$, Chunyi Zhou$^{1}$, Zhou Feng$^{1}$, Songze Li$^{3}$, Shouling Ji$^{1}$}\\
Zhejiang University$^{1}$, Chongqing University$^{2}$, Southeast University$^{3}$\\
\small{$^{\dagger}$Equal contribution}
}

% \author{Anonymous Submission}
\maketitle
\thispagestyle{empty}

% 关键是写之前先逐个回答下面的问题。
% (1) 我们的technical contribution是什么。
% (2) 我们的pipeline解决了什么technical challenge。
% (3) 我们方法本质上能work的原因是什么。
% (4) 我们方法的technical advantage是什么，我们的新认知是什么（重要）。

\begin{abstract}
Federated learning (FL) has attracted substantial attention in both academia and industry, yet its practical security posture remains poorly understood. In particular, a large body of poisoning research is evaluated under idealized assumptions about attacker participation, client homogeneity, and success metrics, which can substantially distort how security risks are perceived in deployed FL systems.
This paper revisits FL security from a measurement perspective. We systematize three major sources of mismatch between research and practice: unrealistic poisoning threat models, the omission of hybrid heterogeneity, and incomplete metrics that overemphasize peak attack success while ignoring stability and utility cost. To study these gaps, we build \textit{\textbf{TFLlib}}, a uniform evaluation framework that supports image, text, and tabular FL tasks and re-implements representative poisoning attacks under practical settings.
Our empirical study shows that idealized evaluation often overstates poisoning risk. Under practical settings, attack performance becomes markedly more dataset-dependent and unstable, and several attacks that appear consistently strong in idealized FL lose effectiveness or incur clear benign-task degradation once practical constraints are enforced. These findings further show that final-round attack success alone is insufficient for security assessment; practical measurement must jointly consider effectiveness, temporal stability, and collateral utility loss.
Overall, this work argues that many conclusions in the FL poisoning literature are not directly transferable to real deployments. By tightening the threat model and using measurement protocols aligned with practice, we provide a more realistic view of the security risks faced by contemporary FL systems and distill concrete guidance for future FL security evaluation.
Our code is availiable at \url{https://github.com/xaddwell/TFLlib}
\end{abstract}

\section{Introduction}
\label{sec:introduction}

In recent years, Federated Learning (FL) has emerged as an important machine learning paradigm for collaborative model training without centralizing raw data~\cite{YangLCT19Federated}. Such a decentralized approach enables multiple entities or devices to contribute to a shared model while keeping training data local, making FL attractive for deployment in healthcare~\cite{NguyenPPDSLDH23Federated}, finance~\cite{LongT0Z20Federated}, and Internet-of-Things (IoT) networks~\cite{NguyenDPSLP21Federated}. As FL moves from research prototypes to production systems, however, its security properties increasingly depend on how the learning process behaves under practical deployment constraints rather than in idealized simulations.

\textbf{The literature on security risks of FL:} Despite its compelling promise, FL faces significant hurdles in its practical implementation and broader adoption due to security risks that undermine model integrity and system availability~\cite{QammarDN22Federated}. Recent works have proposed various attacks and defenses to evaluate and mitigate these risks. Byzantine attacks~\cite{LiNV23Byzantine,XuHSL22Byzantine,ShejwalkarH21Manipulating,FangCJG20Local,MozaffariCH24Fake} compromise partial clients and disrupt the learning process by sending incorrect or maliciously crafted updates to the server, leading to a breakdown in system availability. Backdoor attacks~\cite{xie2019dba,LiYHLWFS233DFed,BagdasaryanVHES20How,WangSRVASLP20Attack,ZhangJCLW23A3FL} attempt to implant targeted misbehavior into the trained model through carefully crafted updates, thereby compromising model integrity. Consequently, researchers have proposed various techniques to protect FL against such attacks, e.g., robust aggregation~\cite{BlanchardMGS17Machine,YinCRB18Byzantine} and anomaly detection~\cite{Luis19Byzantine,MhamdiGR18The}, most of which have not been adopted by FL practitioners (Tab~\ref{tab:practicefl}) due to their complexity in implementation, high computational costs, or the need for additional infrastructure.

\textbf{The gaps from research to practice:} Paradoxically, while the academic landscape is replete with studies proposing novel attacks and defenses, numerous investigations reveal a disturbing disconnect when applied to real-world scenarios~\cite{shejwalkar2022back,khan2023pitfalls,ApruzzeseADFPR23Real,fedscale}. 
Many existing works are anchored in simplified and idealized settings~\cite{ApruzzeseADFPR23Real,khan2023pitfalls}, overlooking the complexities inherent in practical FL ecosystems, such as the hybrid nature of data and system heterogeneity~\cite{ImteajTWLA22A}, and often discount the practicality (e.g., computation and memory cost)~\cite{fedscale}. Furthermore, the prevailing evaluation metrics~\cite{khan2023pitfalls} frequently fall short in capturing the nuanced balance between security robustness, and the overall effectiveness of FL frameworks~\cite{shejwalkar2022back} (Sec~\ref{sec:system}). Consequently, commercial federated frameworks rarely take these research findings and proposed methods into consideration~\cite{unifed}. There seem to be gaps between these studies and real-world scenarios, yet few papers have investigated this disparity. This work, therefore, positions itself at the forefront of addressing these challenges and bridging the gaps between theoretical research and practical deployment considerations in FL. To this end, we endeavor to answer the following pivotal research questions:

$\textbf{\textit{RQ1:}}$ How do existing gaps affect the assessment of FL's security risks between academia and industry? 

$\textbf{\textit{RQ2:}}$ What factors have resulted in the misrepresentations of the assessment, empirically and theoretically? 

$\textbf{\textit{RQ3:}}$ What insights can be gained considering the identified gaps for both FL researchers and practitioners? 

\textbf{Our solutions:} To tackle these questions, our research embarks on a comprehensive measurement framework, which includes the systemization of research gaps, the development of a uniform evaluation framework (TFLlib, \ul{T}rustworthy \ul{F}ederated \ul{L}earning \ul{Lib}rary), the execution of comprehensive experiments across diverse datasets and configurations, the formulation of enhanced performance metrics that encapsulate the complexity of the practical condition. In doing so, our goal is to acknowledge the disparities and propose actionable steps forward to mitigate them.

\textbf{Our contributions:} In this work, we dissect the FL poisoning-security literature and identify crucial disparities between research proposals and practical deployments. Intending to bridge these gaps, we quantify how they distort our understanding of FL robustness in the wild. Our main contributions are as follows:

\paragraphbe{I. Systematic identification of gaps.} We undertake a meticulous analysis that sheds light on the disparities in threat modeling (Sec~\ref{sec:gap1}), overlooked hybrid heterogeneity (Sec~\ref{sec:gap2}), as well as the limitations of evaluation metrics (Sec~\ref{sec:gap3}), thus exposing the disconnect between academic assumptions and the practical realities encountered in FL security research.

\paragraphbe{II. Framework with practical measurement.} To redress and measure these inadequacies, we introduce \textbf{TFLlib} (Sec~\ref{sec:system}), a comprehensive FL framework designed to reassess the efficacy of mainstream poisoning attacks under a more pragmatic lens. This includes re-implementing and evaluating byzantine and backdoor attacks across image, text, and tabular domains. Further, we devise a new suite of performance metrics that captures the multifaceted dimensions of practical security evaluation.
% These metrics cater to the nuanced complexities inherent in FL, ensuring a thorough appraisal of FL risks in practical contexts.

\paragraphbe{III. Practical insights and theoretical analysis.} Leveraging a broad spectrum of datasets, we test the robustness of prevalent FL scenarios under our refined metrics. By incorporating data and system heterogeneity, we ensure that evaluations better mirror real deployments. Our findings yield actionable insights into the actual security risks confronting contemporary FL frameworks. We outline recommendations for improving how future FL poisoning attacks and defenses should be evaluated under practical constraints.

\paragraphb{Implications of our study:} Collectively, our measurement shows that practical FL security evaluation must be grounded in realistic attacker participation, hybrid heterogeneity, and utility-aware metrics. We hope this work sharpens the community's understanding of poisoning threats in deployed FL systems and helps build more reliable and security-conscious FL ecosystems.

\section{Preliminary}
\label{sec:preliminary}

\subsection{Federated Learning}
Federated learning, first introduced by Google~\cite{mcmahan2017communication}, is a decentralized learning framework designed to train machine learning models across multiple devices, holding data locally without data centralization. There are cross-device and cross-silo horizontal FL tasks with different scales, controls, and types of participants involved~\cite{YangWXCBLL21Characterizing}. In this paper, we focus on cross-device and asynchronous settings since it has been widely deployed in real-world scenarios~\cite{bonawitz2019towards,fate,huba2022papaya}. Cross-device FL primarily focuses on scenarios with a large number of consumer devices, such as smartphones or IoT devices, participating in the learning process. The emphasis here is on scalability, dealing with highly non-independent and identically distributed (Non-IID) data, and ensuring efficient communication given the constraints of mobile networks and device capabilities.

The core principle of FL operates on a client-server architecture. In this setup, for each selected client $c_{i}\in\{c_1,c_2,..c_N\}$, where $i\in\{1,2,..K\}$ and $K$ denote the number of the clients selected for aggregation (join ratio $\alpha=\frac{K}{N}$), they retain private data $\mathcal{D}_i$ locally and perform a set of iterations of training for local step $j\in\{1,2,...E_i\}$ at round $t\in\{1,2,..T\}$:
\begin{equation}
    W_{i, t}^{j+1}=W_{i, t}^{j}-\eta\cdot\nabla\mathcal{L}(W_{i, t}^{j},\mathcal{D}_i)
\end{equation}
where $\nabla\mathcal{L}(\cdot)$ is the gradient of $W_{i, t}^{j}$ for each iteration of local training and $W_{i, t}$ is then updated locally with learning rate $\eta$ (e.g., using Stochastic Gradient Descent~\cite{bottou2010large}). Note that $W_{i, t}^{0}$ is set to $G_{t-1}$ (global model at round $t-1$) before local training at round $t$. Therefore, the local update of model parameters is $\nabla_{i,t} = W_{i, t}^{S_i}-G_{t-1}$. These updates, rather than the actual data, are then transmitted to the central server. Without loss of generality, the server can aggregate these updates with FedAvg~\cite{mcmahan2017communication} to refine a global model with weight $w_{i}$:
\begin{equation}
G_{t}=G_{t-1}+\sum_{i=1}^{K}w_{i}\nabla_{i,t}
\end{equation}
where $G_{t}$ denotes the aggregated model at round $t$ and $w_{i}$ is calculated via the ratio of the data at round $t$: $w_{i}=\frac{\|\mathcal{D}_i\|}{\sum_{i=1}^{K}\|\mathcal{D}_i\|}$. For simplicity but without the loss of generality, we select FedAvg~\cite{mcmahan2017communication} as the aggregation algorithm in this paper since it's widely applied by FL platform in both industry and academia.

Note that each communication round has a fixed length of wall-clock time; thus, straggling clients who take too much time for training models (device heterogeneity) or uploading local updates (communication heterogeneity) are not selected. Considering the diversity of device resources~\cite{Zeng23Tackling}, different clients typically adopt different epochs $E_i$ of local training according to the amount of data $\|\mathcal{D}_i\|$. Intuitively, we can denote this relation as $E_i\propto \frac{1}{\|\mathcal{D}_i\|}$ and $E_i\propto \tau_i^{d}$, where $\tau_i^{d}\in[0,1]$ stands for the device condition of client $i$ and higher $\tau_i^{d}$ means less computation resources. Besides, communication heterogeneity~\cite{bonawitz2019towards} represents that the communication conditions of different clients vary greatly, some of which may be disconnected before aggregation. Similarly, we denote the communication condition of client $i$ as $\tau_i^{c}\in[0,1]$ and lower $\tau_i^{c}$ means less possibility of encountering disconnection. For simplicity, here we assume that these two variables ($\tau_i^{d}$ and $\tau_i^{c}$) are independent and unrelated. Note that the detailed implementation of these heterogeneities is discussed in Sec~\ref{sec:gap2}.

Once the server receives all local updates, aggregation ensues, followed by directives issued to the contributing devices regarding their next connection schedule. Subsequently, model $G_{t}$ is disseminated back to the clients. This iterative process continues until the model converges while keeping training data decentralized across participants.

\begin{table*}[ht]
\tabcolsep=0.15cm
\renewcommand{\arraystretch}{1.2}
\scriptsize
% \vspace*{-.5em}
\centering
\caption{FL platforms from industry and academia, with corresponding functionality, aggregation algorithms, privacy-preserving techniques, and security guarantees. \textit{N/A} means that this part is not supported.}
\label{tab:practicefl}
\begin{tabular}{c|cccccc}
\toprule
\textbf{Source} & \textbf{FL Platform} & \textbf{Institute} & \textbf{Functionality} & \textbf{Aggregation Algorithm} & \textbf{Privacy Preserve} & \textbf{Security Guarantee} \\ 
\midrule
\midrule
\multirow{9}{*}{\textbf{Industry}} & TFF~\cite{tff} & Google & Simulator & FedAvg, FedProx & \cellcolor{green!10}\textit{DP} & \cellcolor{red!30}\textit{N/A} \\
 & FLUTE~\cite{garcia2022flute} & Microsoft & Simulator & FedAvg, DAG, etc & \cellcolor{green!10}\textit{DP} & \cellcolor{red!30}\textit{N/A} \\
 & FATE~\cite{fate} & WeBank & Framework & FedAvg & \cellcolor{green!30}\textit{HE, MPC, SA} & \cellcolor{red!30}\textit{N/A} \\
 & IBM FL~\cite{ibmfl} & IBM & Framework & FedAvg, FedProx, etc & \cellcolor{green!40}\textit{DP, HE, MPC, SA} & \cellcolor{green!10}\textit{Krum} \\
 & MindsporeFL~\cite{mindsporefl} & Huawei & Framework & FedAvg & \cellcolor{green!40}\textit{DP, MPC, SA, TEE} & \cellcolor{red!30}\textit{N/A} \\
 & PySyft~\cite{pysyft} & OpenMined & Framework & FedAvg & \cellcolor{green!30}\textit{DP, HE, MPC} & \cellcolor{red!30}\textit{N/A} \\
 & FedLearner~\cite{fedlearner} & ByteDance & Framework & FedAvg & \cellcolor{green!20}\textit{HE, MPC} & \cellcolor{red!30}\textit{N/A} \\
 & FedScope~\cite{fedscope} & Alibaba & Simulator, Framework & FedAvg, Ditto, etc  & \cellcolor{green!30}\textit{DP, HE} & \cellcolor{green!40}\textit{Bulyan, Krum, etc} \\
 & PaddleFL~\cite{paddlefl} & Baidu & Simulator, Framework & FedAvg & \cellcolor{green!40}\textit{DP, HE, MPC, SA} & \cellcolor{red!30}\textit{N/A} \\ 
 \midrule
\multirow{5}{*}{\textbf{Academia}} & FedScale~\cite{fedscale} & UMich & Simulator & FedAvg, FedYoGi, etc & \cellcolor{green!20}\textit{DP, SA} & \cellcolor{red!30}\textit{N/A} \\
 & FedML~\cite{fedml} & USC & Simulator, Framework & FedAvg, FedNova, etc & \cellcolor{green!30}\textit{DP, MPC, SA} & \cellcolor{red!30}\textit{N/A} \\
 & FedLab~\cite{fedlab} & UESTC & Simulator & FedAvg, SCAFFOLD, etc & \cellcolor{green!10}\textit{DP} & \cellcolor{red!30}\textit{N/A} \\
 & PFLlib~\cite{zhang2023pfllib} & SJTU & Simulator & FedAvg, SCAFFOLD, etc & \cellcolor{green!10}\textit{DP} & \cellcolor{red!30}\textit{N/A} \\
 & Flower~\cite{flower} & Cantab & Simulator & FedAvg, FedProx, etc & \cellcolor{green!20}DP, SA & \cellcolor{green!20}\textit{Bulyan, Krum} \\
\bottomrule
\end{tabular}
\begin{tablenotes}
\footnotesize
\item[1] \textbf{Abbreviation}: DP: Differential Privacy; HE: Homomorphic Encryption; MPC: Multi-Party Computation; SA: Secure Aggregation; TEE: Trusted Execution Environment. ``Simulator" means that the platform is used for FL execution simulation on one or multiple devices, while ``Framework'' denotes that the platform provides users with service for real-world deployment and applications.
\end{tablenotes}
\end{table*}

\subsection{Security Risks of Federated Learning}
Training without sharing data is one of the main motivations for applying FL. However, FL also introduces significant vulnerabilities that malicious adversaries can exploit through the interaction between the server and clients. In this paper, we focus on poisoning attacks, where untrusted clients craft malicious updates to threaten the integrity~\cite{ZhangJCLW23A3FL,BagdasaryanVHES20How} and availability~\cite{ShejwalkarH21Manipulating,BaruchBG19A} of the collaborative learning process and the global model. We leave privacy attacks and privacy-preserving countermeasures to future work, as discussed in Sec.~\ref{sec:future_work}.

\subsubsection{Poisoning Attack Against Federated Learning} 
Poison attacks in FL aim to disrupt or manipulate the learning process by introducing adversarial updates $\nabla_{i,t}^{adv}$ during global aggregation. From the adversaries' perspective, poisoning local updates is the only way to achieve this. Formally, we define the unified poison attack model as follows:
\begin{equation}
\nabla_{i,t}^{adv} = \mathcal{F}_{\epsilon}(G_{t},\mathcal{L}_{adv}(\mathcal{T},\mathcal{D}_{i}),\mathcal{I}_{aux})
\label{eq:poison_eq}
\end{equation}
where the different notations define the adversary’s capability, knowledge, and strategy. By poisoning the local dataset $\mathcal{D}_{i}$ with function $\mathcal{T}$, adversaries can inject malicious backdoor into the global model, e.g., for $(\hat{x},y_t)\in\mathcal{T}(D_{i})$, $\hat{x}$ and $y_t$ represents the triggered samples and its target class respectively. The auxiliary information $\mathcal{I}_{aux}$ (e.g., extra datasets, other clients' updates, etc.) represents the knowledge of the adversary and empowers them to act stealthily and maximize the impact of the attacks. With the above knowledge, the attacker leverages strategy $\mathcal{L}_{adv}$ to optimize the malicious updates. To maintain a balance between attack effectiveness and stealthiness, adversaries adopt constraint $\mathcal{F}_{\epsilon}$ to bound updates. Specifically, scaling and $\epsilon$-constraint are commonly used transformations to evade detection. By manipulating local updates with $\mathcal{F}_{\epsilon},\mathcal{T}$ and $\mathcal{I}_{aux}$, adversaries can poison the global models iteratively. As for different attack objectives, two primary categories exist primarily: byzantine and backdoor attack, designed to degrade model performance and introduce targeted misclassification, respectively. 

\textbf{Byzantine attacks (BZAs)}, also known as untargeted poison attack~\cite{LiNV23Byzantine,YinCRB18Byzantine,Luis19Byzantine,XuHSL22Byzantine,FangCJG20Local,ShejwalkarH21Manipulating,cao2022mpaf}, aims to inject uncertainty, inconsistency, or dysfunction into the system by modifying the local updates with $\mathcal{F}_{\epsilon}(\cdot)$, reducing availability of FL. Additionally, if the global model is disrupted, participants may lose confidence and withdraw, further impacting the system's operational integrity.

\textbf{Backdoor attacks (BKAs)} focus on surreptitiously implanting a hidden functionality ``backdoor'' into the globally trained model~\cite{LiYHLWFS233DFed,xie2019dba,LyuHWLWL023Poisoning,ZhangJCLW23A3FL,ZhangPSYMMR022Neurotoxin,WangSRVASLP20Attack,BagdasaryanVHES20How} by poisoning the local dataset $D_{i}$ with transformation $\mathcal{T}(\cdot)$. This backdoor can later be triggered to manipulate the model's behavior on specific inputs while maintaining normal or high accuracy on regular data. This corruption breaches the trust in the model's outputs and decisions, demonstrating a clear violation of integrity.

\subsection{Federated Learning in Practice}
\label{sec:fl_in_practice}
Since 2017, when Webank introduced the first industrial-grade open-source FL framework, FATE~\cite{fate}, a series of major domestic and international internet corporations, including Huawei~\cite{mindsporefl}, Alibaba~\cite{fedscope}, Google~\cite{tff}, and OpenMinded~\cite{pysyft}, are joining this track. FL has gradually transitioned from academic research to commercial products in the industry. As shown in Tab~\ref{tab:practicefl}, there have been many FL frameworks (for real-world applications) and simulators (for FL simulations) from industry and academia. We compile the aggregation algorithms and security guarantees adopted by these platforms. Unexpectedly, even though there has been much research about poison-robust FL~\cite{SandeepaSWL24SHERPA,CaoF0G21FLTrust,LycklamaBVKH23RoFL}, most of the listed platforms still rely on simple aggregation and offer limited explicit security guarantees. This raises an important question: \textit{\textbf{Why is there such a significant disparity between the research on secure FL in academia and the practical choices made by these industry platforms? What factors are driving this disparity?}} This motivates us to delve deeper into the exploration of the existing gaps between research and practice in the following sections on threat model, evaluation, and metrics.
\begin{takeaway}
\textbf{Motivation.} Most existing commercial FL platforms still adopt simple and efficiency-oriented security designs rather than the more sophisticated defenses proposed by recent research, exhibiting a large disparity between academia and industry.
\end{takeaway}

% The commericial FL framework is designed to provide users with easy-to-use APIs, flexible and extendable module implementations and compliant communication and scheduler management. These features enable users to effortlessly integrate their own components, including datasets, customized model structure, aggregation algorithms, evaluation metrics, etc. 

% The framework usually consists of three parts: \textbf{communication management, scheduler and aggregator.} The communication framework ensures that the server remains connected to each client, and exchange model parameters and other information through a secure channel. Scheduler collects server information during cluster initialization and ensures cluster consistency. During the training process, the scheduler manages each client in real-time to ensure the client's dynamic entry and exit. At the same time, the efficiency of node communication is improved through parallel and pipeline scheduling. Finally, the aggregator selects several clients in each round, delivers model parameters to them, collects updates, and aggregates model parameters for the next round. Some FL frameworks also provide cloud computing platforms and offer computing power to users. Meanwhile, considering the privacy protection issue, the framework often provides additional defense mechanisms, such as SSL security authentication in the communication process, differential privacy protection, homomorphic encryption and secure multi-party computation in the training process.

\section{Security Gaps in the Wild}
\label{sec:motivation}

In this section, we discuss the existing gaps between research and practices from three dimensions: \textit{1) threat models, 2) evaluation, and 3) metrics}. We argue that many of the existing research studies fail to present a representative and comprehensive evaluation under practical threat models. In contrast, our work aims to systematically analyze how these gaps affect the measurement of security risks of FL and propose actionable strategies to bridge these gaps in the next section.

\subsection{Threat Model (Gap \uppercase\expandafter{\romannumeral1})}
\label{sec:gap1}
\label{sec:threat_model_poison}
\textit{(1) Adversary's objective.} Inspired by~\cite{shejwalkar2022back}, poisoning attacks seek to violate the integrity and availability of the FL system. Specifically, these objectives can be categorized as follows. \\
\textit{\textbf{Attack specificity.}} The attack aims to manipulate the model's behavior on specific input~\cite{} (e.g., with trigger) via $\nabla_{i,t}^{adv}$. \\
\textit{\textbf{Utility degradation.}} The attack intends to degrade the performance~\cite{cao2022mpaf,LiNV23Byzantine,XuHSL22Byzantine} of the global model, disrupting the service for legitimate clients. \\
\textit{\textbf{Poison stealthiness.}} In the era of the arms race, successful poison means achieve effectiveness while avoiding potential detection of $\nabla_{i,t}^{adv}$~\cite{} from the server, using $\mathcal{F}_{\epsilon}$.

\textit{(2) Adversary's knowledge.} For FL service, launching poisoning attacks from the client's side means that there have been compromised clients who act as attackers with $\mathcal{I}_{aux}$. \\
\textit{\textbf{Restricted access.}} As illustrated in Eq~\ref{eq:poison_eq}, attackers share the same knowledge (e.g., $G_t, \mathcal{D}_i$) with compromised clients, a well-grounded assumption in empirical evidence, where the adversary can control a small number of clients. \\
\textit{\textbf{Exterior access.}} However, some byzantine attacks~\cite{ShejwalkarH21Manipulating,PoisonedFL} require access to $\nabla_{i,t}$ from benign clients to generate $\nabla_{i,t}^{adv}$, which we argue that it's impractical since the training time of the malicious updates depends on the straggler among them (benign and malicious), raising the risk of upload failure.

\textit{(3) Adversary's capability.} Next, we elaborate on the adversary’s capability to achieve his objective. \\
\textit{\textbf{Aggregation participation.}} Only the selected malicious clients can eventually poison $G_t$ through $\nabla_{i,t}$. (\uppercase\expandafter{\romannumeral1}) \textit{Fixed participation}: Previous works~\cite{BagdasaryanVHES20How,ZhangPSYMMR022Neurotoxin,xie2019dba,LyuHWLWL023Poisoning} tend to assume a fixed number (percentage) of attackers to participate in the aggregation, significantly enlarging the capability of the adversary in practical scenarios. Actually, the server randomly selects the active clients, meaning that the number of times an attacker can participate in aggregation is far less than the previous assumption. (\uppercase\expandafter{\romannumeral2}) \textit{High malicious percentages}: Other than fixed participation, previous attacks also require high percentages of malicious clients (e.g., 20\%-50\%) to remain effective~\cite{MozaffariCH24Fake}. Since the cost of manipulating a compromised client botnet at scale (which includes breaking into devices~\cite{shejwalkar2022back}) is non-trivial, a small percentage of compromised clients is more in line with the practice. \\
\texttt{\textbf{Case analysis.}} Given an FL simulation with 100 clients, 10 of which are selected randomly for aggregation each round, launching an attack with no less than 4 malicious clients selected each round (99\% probability) means \textbf{at least 71 of the total 100 clients should be compromised!} \\
\textit{\textbf{Extra computation.}} Considering the FL scenarios where the computational resource varies among clients, several poisoning attacks introduce additional computation overhead, increasing the time malicious clients spend on computing malicious updates $\nabla_{i,t}^{adv}$ before uploading them. Consequently, the adversary requires extra computation for generating $\nabla_{i,t}^{adv}$. \\
\textit{\textbf{Attack frequency.}} To amplify the performance of the attack, previous attacks~\cite{xie2019dba,BagdasaryanVHES20How,WangSRVASLP20Attack,LyuHWLWL023Poisoning} also assume that malicious clients can launch attacks at specific rounds and last for several rounds. However, considering that each client is selected randomly, such an assumption grossly overestimates the adversary's capabilities, inconsistent with real-world situations. \\
\texttt{\textbf{Case analysis.}} Given an FL simulation with the join ratio $\alpha=0.1$, that is, 10\% of clients are randomly selected for aggregation each round. Launching an attack lasting for $5$ round means that \textbf{there's a $\alpha^{5}=1e-5$ (0.001\%) chance of it happening, almost impossible!} \\
\textit{\textbf{Weight scaling.}} Many existing works~\cite{xie2019dba,BagdasaryanVHES20How} also leverage weight scaling to enlarge the magnitude of the malicious updates $\nabla^{adv}_{i,t}$ since they assume that FL employs secure aggregation to protect the confidentiality of participants' local updates and thus cannot detect the malicious clients~\cite{BagdasaryanVHES20How}. However, our investigation in Tab~\ref{tab:practicefl} demonstrates that more than half of the industrial platforms do not employ secure aggregation~\cite{MansouriOJC23SoK}. Moreover, our experimental results also reveal that weight scaling can significantly affect the benign accuracy of the FL tasks, and thus, practical attackers should never use weight scaling operations for stealthiness.
\begin{takeaway}
\textbf{Gap I (Security).} Existing research on poisoning attacks in FL often relies on unrealistic assumptions about threat models, which misrepresents the practical effectiveness of the attack. Previous studies typically assume high percentages of malicious clients, fixed participation in model aggregation, and exterior access to benign clients' updates, scenarios that are rarely feasible in real-world FL systems.
\end{takeaway}
\vspace{5pt}
\ul{\textbf{What we argue to be practical.}} In practice, attacks are constrained by the random selection of clients, limited numbers of compromised participants, and the absence of weight scaling, which might significantly reduce the adversary’s impact. To bridge these gaps, our work advocates for more realistic and practical threat models that align with real-world FL systems, ensuring a more accurate assessment of security risks. Expressly, we assume that malicious clients can only obtain historical updates from each other. By assuming a fixed number of compromised clients or an unrealistically high percentage of malicious clients' participation, the effectiveness of previous poisoning attacks might be misrepresented. Instead, we adopt real-world FL schemes~\cite{LiYHLWFS233DFed,ZhangJCLW23A3FL}, selecting clients randomly among only $1\%-10\%$ malicious clients of the participants with no weight scaling.

\newcolumntype{L}[1]{>{\raggedright\arraybackslash}p{#1}}

\begin{table*}[t]
\centering
\scriptsize
\setlength{\tabcolsep}{0.12cm}
\renewcommand{\arraystretch}{1.22}
\caption{Key gaps between prior FL security evaluation and the practical poisoning-security setting studied in this paper.}
\label{tab:gaps}
\begin{tabular}{L{1.75cm} L{1.9cm} L{3.35cm} L{4.35cm} L{4.45cm}}
\toprule
\textbf{Gap} & \textbf{Aspect} & \textbf{Common in prior work} & \textbf{What we argue is practical} & \textbf{Why it matters} \\
\midrule
\multirow{14}{*}{\textbf{Threat Model}}
& Adversary knowledge
& Some attacks assume access to benign clients' updates when crafting malicious updates~\cite{ShejwalkarH21Manipulating}.
& Malicious clients should only rely on their own local state and historical information obtainable from other compromised clients.
& Benign-update access substantially strengthens the attacker and can make poisoning appear feasible even when the assumption is unrealistic in deployed FL. \\

& Aggregation participation
& Fixed attacker participation across rounds and persistent multi-round attacks are often assumed~\cite{xie2019dba,BagdasaryanVHES20How,WangSRVASLP20Attack,LyuHWLWL023Poisoning}.
& Client participation should be random, with attack opportunities determined by the actual join ratio rather than a fixed attacker schedule.
& Random participation sharply reduces the frequency and coordination of poisoning opportunities, which lowers practical attack strength. \\

& Malicious population
& High malicious ratios, often far above what is plausible in production deployments, are used to keep attacks effective~\cite{MozaffariCH24Fake}.
& The attacker should control only a small fraction of clients (e.g., 1\%--10\%) and still compete under random sampling.
& Attack success under large compromised populations can overestimate the real deployment risk by assuming an attacker budget that is economically implausible. \\

& Stealth mechanism
& Weight scaling and other aggressive update manipulations are frequently used to force success~\cite{xie2019dba,BagdasaryanVHES20How}.
& Practical attackers should be evaluated without assuming secure aggregation is always present and without relying on obviously utility-damaging scaling tricks.
& A threat model that ignores benign-task degradation can incorrectly label conspicuous attacks as ``stealthy.'' \\
\midrule
\multirow{16}{*}{\textbf{Evaluation}}
& Aggregation and defense setup
& Prior evaluations use heterogeneous FL pipelines, including FedSGD-style settings and sophisticated robust defenses that are uncommon in practice~\cite{khan2023pitfalls}.
& Evaluation should default to representative deployment choices such as FedAvg and should distinguish clean measurement from defense-specific benchmarking.
& Inconsistent pipelines make attack numbers incomparable and can invert conclusions about which attacks or defenses matter in practice. \\

& Data and modality scope
& Many studies benchmark mainly on simple image datasets and narrow task families~\cite{khan2023pitfalls}.
& Practical evaluation should cover image, text, and tabular FL workloads with corresponding model families.
& Security conclusions drawn from one easy modality may not transfer to real systems that span different data types and learning dynamics. \\

& Model architecture scope
& CNN-centric evaluation is common even when production FL uses diverse backbones.
& Benchmarks should include MLPs, CNNs, and Transformers whenever the target application mix requires them.
& Attack behavior can be architecture-dependent, so single-family evaluation risks reporting only a narrow slice of the threat landscape. \\

& Hybrid heterogeneity
& IID data, homogeneous devices, and reliable communication are often treated as default assumptions~\cite{fedscale,leaf,YangWXCBLL21Characterizing}.
& Statistical, device, and communication heterogeneity should be modeled jointly.
& Heterogeneity changes convergence, participation, and timing, all of which directly affect whether poisoning remains effective outside idealized simulations. \\
\midrule
\multirow{10}{*}{\textbf{Metrics}}
& Attack specificity
& Backdoor effectiveness is often summarized by a single final-round ASR.
& Practical backdoor evaluation should report sustained specificity and stability, e.g., BSA and BSV over the converged training tail.
& Peak or last-round success can exaggerate attacks that are unstable or only succeed in a narrow operating regime. \\

& Utility degradation
& Byzantine impact is commonly reported with one-shot or coarse accuracy-drop summaries.
& Utility degradation should be evaluated over the stable tail of training, e.g., with BDA and BDV.
& Practical disruption is a temporal phenomenon; single-point reporting can miss whether degradation is durable or merely transient. \\

& Poison stealthiness
& Stealthiness is often inferred only from whether an attack bypasses a given defense.
& Stealthiness should also account for benign-task utility and temporal variance, such as ACC and ACCV.
& An attack that triggers reliably but visibly harms the main task is a different operational risk from one that remains both effective and quiet. \\
\bottomrule
\end{tabular}
\end{table*}

\subsection{Evaluation Discrepancy (Gap \uppercase\expandafter{\romannumeral2})}
\label{sec:gap2}
Considering the evaluation of FL for measuring security risks, we also find that previous works tend to adopt various FL configurations (e.g., aggregation algorithm, evaluation scope, etc), offering even inconsistent results~\cite{KhanSHA23On}. Here, we figure it out and provide unified and practical configurations adopted in this paper.

\textit{(1) Various FL setups.} A unified and representative setup is necessary to simulate the practical FL runtime and offer generalizable insights for practitioners. For instance, the joining ratio, the number of participants, aggregation algorithms, etc, are critical to evaluate FL. Without loss of generality, we list the primarily related parts below.  \\
\textit{\textbf{Aggregation algorithm.}} As shown in Tab~\ref{tab:practicefl}, most companies use FedAvg~\cite{mcmahan2017communication} as the aggregation algorithm instead of others (e.g., FedSGD~\cite{mcmahan2017communication}) for its efficiency and popularization. As revealed by ~\cite{khan2023pitfalls}, FedAvg is less susceptible to poisoning attacks since its faster convergence, leaving less possibility for an adversary to launch poisoning. Surprisingly, there are still about 40\% of prior works~\cite{khan2023pitfalls} use FedSGD-based aggregation with slower convergence and higher communication for evaluations. \\
\textit{\textbf{Poisoning-robust defenses (PRDs).}} The practical FL task often prioritizes efficiency and simplicity, adopting less complex defense mechanisms. In contrast, previous attack studies mainly consider sophisticated and idealized defenses proposed in research settings~\cite{shejwalkar2022back}. In comparison, Tab~\ref{tab:practicefl} also illustrates that most of the FL platforms choose none SOTA defense methods proposed by academia. \\
\textit{\textbf{Others.}} Previous studies tend to use different experimental parameters (e.g., the number of FL participants, joining ratio), which introduced great obstacles to benchmarking different FL optimizations and even misled the evaluation. Therefore, we tried to choose more representative parameters, which we suggest that future work can also align with as much as possible to facilitate better horizontal comparison.

\textit{(2) Unitary evaluation scope.} In practical settings, FL tasks often deal with diverse data types such as electronic health records, financial transactions, or textual communications, including text, tabular and image data~\cite{fedscale}. However, previous attacks~\cite{} or defenses~\cite{cao2023fedrecover} mainly focus on datasets like MNIST or FashionMNIST~\cite{khan2023pitfalls} for evaluation, which means that these limited scopes may lose the generalizable risk representation scenarios and fail to capture the complexity and variety of data encountered in practical FL. \\
\textit{\textbf{Unitary data types.}} Previous works often focus on a single data type (i.e., image data). However, in real-world FL applications, multiple data types might be involved. For example, a healthcare FL system could handle patient medical structured tabular data~\cite{KaissisMRB20Secure}. Failure to do so can result in security mechanisms that are ill-equipped to handle real-world data diversity, leading to potential vulnerabilities and inefficiencies when these mechanisms are deployed in practice. \\
\textit{\textbf{Unitary data complexity.}} Many existing studies evaluate datasets with relatively simple data structures and low complexity~\cite{khan2023pitfalls}. However, data can be highly complex, with interrelated features and hierarchical structures in practical FL~\cite{fedscope}. For instance, financial transaction data may have complex temporal patterns and correlations. Ignoring such complexity leads to an incomplete understanding of the security risks~\cite{khan2023pitfalls}. \\
\texttt{\textbf{Case analysis.}} As demonstrated by Khan et al.~\cite{khan2023pitfalls}, strategies like FedRecover rated that FedRecover excels on balanced datasets like MNIST and FashionMNIST but \textbf{falters with more complex and diverse datasets} (e.g., CIFAR10). This discrepancy highlights the importance of considering data diversity in evaluating FL security risks. \\
\textit{\textbf{Diverse model architectures.}} Previous works mainly use CNNs as the FL backbone model, which narrows the evaluation scope because different data types in practice typically entail different architectures (i.e., MLPs, CNNs, and Transformers for tabular, image, and text data, respectively). A security evaluation restricted to one model family can therefore miss important attack behaviors that only emerge in more heterogeneous deployments. \\
% \textbf{3) Metrics for security risks}: There is a dearth of metrics designed to measure the security risks of these practical frameworks accurately, leaving the true extent of risks largely unexplored and unaddressed. This gap in measurement means that stakeholders may be unaware of the actual vulnerabilities and the security posture of their deployed FL systems.
% This gap leads to whether these practical deployments are more vulnerable than anticipated, as they do not incorporate these sophisticated defenses, exposing them to potential security risks. \textbf{2) Protocol disparities}: The communication protocols and overall operational settings of practical FL frameworks (such as PySyft, FedML, and Mindspore) are vastly different from those considered in research. Practical frameworks often optimize for reduced communication time, efficient bandwidth usage, and dynamic client selection, while research settings might assume idealized, static environments with stable communication channels. These discrepancies mean that academic evaluations may overlook the complexities and constraints faced in real-world applications, leading to ineffective or impractical recommendations. 

\textit{(3) Hybrid heterogeneity.} Unlike the experimental scenarios in which each client has identical computational capabilities~\cite{fedscale}, equal amounts of IID dataset~\cite{leaf}, and operates within a similar computational environment~\cite{YangWXCBLL21Characterizing}, the FL framework in a real-world scenario encounters significantly greater complexity due to heterogeneity, which can be categorized into three types: statistical, device and communication heterogeneity. \\
\textbf{\textit{Statistical heterogeneity.}} According to~\cite{ye2023heterogeneous}, there are generally four types of statistical heterogeneity in FL: label, feature, quality, and quantity skew. In this paper, we focus on the most common and challenging label heterogeneous situation~\cite{LiSBS20Fair,khan2023pitfalls}, Dirichlet distribution~\cite{leaf}. Formally, we use $\tau^{s}\in(0,+\infty)$ to represent the intensity of Dirichlet heterogeneity \textit{Dir($\tau^{s}$)}, and a lower $\tau^{s}$ means a more intense heterogeneity. \\ %As revealed by ~\cite{ChenYJY23APFed},  \\
\texttt{\textbf{Case analysis.}} In the healthcare setting, large hospitals typically treat patients and exhibit lower misdiagnosis rates than small ones, thus with larger and higher quality data. Furthermore, since people with different diseases tend to choose different hospitals, the label and feature distributions may differ significantly. \\
\textbf{\textit{Device heterogeneity.}} For cross-device horizontal FL with a large number of clients, each of them encompasses variations in storage capacity, CPU performance, and network latency~\cite{aibenchmark}. This diversity in device capabilities leads to discrepancies in computation times and can even lead to the failure of certain devices. To address these challenges, different clients implement personalized local training settings~\cite{zhang2023pfllib,fedscale} (e.g., epochs) tailored to their computation resources and data volume to balance computational efficiency and fairness. \\
\texttt{\textbf{Case analysis.}} When using MobileNet-V3 as the backbone model for FL inference ($346\times346$-px image with precision FP16 and batchsize=4), Apple iPhone 16 Pro only need 1.7ms while Samsung Galaxy Tab S8 takes 3ms~\cite{aibenchmark}. For same amount of local data, clients with less computation resources have to provide model updates with deficient training to catch the FL submission.\\
\textbf{\textit{Communication heterogeneity.}} The era of IoT is characterized by a vast spectrum of devices operating under varying network conditions, each with distinct bandwidth capacities, latency, and reliability~\cite{ReisizadehTHMP22,ParkHCM21,YangWXCBLL21Characterizing}. Once devices with low network conditions are selected for aggregation, there is a chance that they will fail to upload their local updates to the server due to poor communication conditions, affecting overall performance and convergence of the global model. This is a practical situation that frequently occurs in real-world applications but has been overlooked in previous works. 

\begin{takeaway}
\textbf{Gap II.} There is a significant discrepancy in how FL systems are evaluated, primarily due to inconsistencies in experimental setups, data types, and real-world complexities. Many studies use unrealistic or overly simplified FL configurations, such as idealized aggregation algorithms, limited data types, and uniform client conditions, leading to inaccurate or non-generalizable results. Additionally, the lack of consideration for heterogeneity in FL, such as variations in client devices, network conditions, and data distributions, often causes substantial differences between research and practical scenarios.
\end{takeaway}
\vspace{5pt}

\ul{\textbf{What we argue to be practical.}} To better align with real-world applications, we embrace diverse data types and complexities and incorporate realistic heterogeneity to provide more accurate assessments. Expressly, we only use FedAvg and adopt no poison-robust defense in order to measure poisoning risks under a simple but realistic deployment scenario. Other detailed configurations are included in Tab~\ref{tab:exp_cfg}. The recent literature typically conducts experiments on CNNs with limited data types and ideal settings. This significant divergence means that many FL attack and defense strategies validated on research datasets may not perform as expected when faced with imbalanced data distributions found in practice. Instead, we expand the scope to tabular, image, and text data on MLP, CNNs, and Transformers. To measure the security risks of FL in the wild, we simulate the hybrid heterogeneity in \textit{TFlib} based on the previous works~\cite{fedscale,zhang2023pfllib,leaf}. The specific implementations are presented in Sec~\ref{sec:system}.

\subsection{Performance Metrics (Gaps \uppercase\expandafter{\romannumeral3})}
\label{sec:gap3}
As mentioned in Sec~\ref{sec:threat_model_poison}, malicious clients have various poisoning objectives, which are formalized in this subsection with proposed metrics. Different metrics, considering the identified gaps, are crucial to capturing and understanding FL's security risks. We first revisit the previous metrics, identify the gaps under the above discrepancy, and then reformulate the metrics, which we argue to be practical. 
% Reflect Real-World Complexity: Traditional metrics often oversimplify FL environments, ignoring the heterogeneity of data and systems, the presence of defenses, and practical constraints. Tailored metrics take these complexities into account, ensuring that assessments mirror the dynamics of actual FL deployments.
% Bridge Theory and Practice: By identifying and addressing the gaps, these metrics help bridge the divide between theoretical research outcomes and practical applicability. They enable researchers to validate theories against practical scenarios, enhancing the reliability and relevance of their findings.
% Comprehensive Risk Assessment: Different attacks pose unique challenges that necessitate distinct evaluation criteria. Poisoning attacks prioritize stealth and durability, while privacy attacks emphasize the amount of information leakage and the ability to evade defenses. Custom metrics provide a nuanced view of each type of risk, leading to a more holistic risk assessment.
% Inform Robust Defense Strategies: Understanding the specific weaknesses and strengths of attacks through targeted metrics guides the development of more effective defenses. Metrics that consider attack stealthiness, for instance, can inspire the creation of advanced detection algorithms that specifically counter stealthy tactics.
\textit{\textbf{Attack specificity.}} For backdoor attacks~\cite{}, previous methods measure the attack specificity by reporting the attack success rate (\textit{ASR}, $asr$) of the last FL training round. Moreover, recent attacks focus on attack durability, i.e., ASR after the attacker stops uploading poisoned updates. However, we emphasize that the specificity of the practical attack should involve the stability of the attack since the backdoor task often conflicts with the main task~\cite{WangSRVASLP20Attack,ZhangPSYMMR022Neurotoxin}, leading to significant fluctuations in ASR and benign accuracy (\textit{BA}, $acc$). Besides, these fluctuations also influence the confidence in the reported performance of the last round. Instead, we propose to use the ASR of the last 10\% rounds (i.e., $\alpha=0.9$) to capture the attack specificity. Specifically, we calculate backdoor specificity accuracy (\textit{BSA}, $\mu_{asr}$) and variance (\textit{BSV}, $\sigma_{asr}$): 
\begin{equation}
    \textbf{\textit{BSA}} = \frac{\sum_{t=\lfloor \alpha\cdot T\rfloor}^{T} asr_{t}}{\lfloor (1-\alpha)\cdot T\rfloor}, \textbf{\textit{BSV}} = \sqrt{\frac{\sum_{t=\lfloor \alpha\cdot T\rfloor}^{T} (\mu_{asr} - asr_{t})^{2}}{\lfloor(1-\alpha)\cdot T\rfloor}}
\label{eq:backdoor_specificity}
\end{equation}
where $asr_t$ denotes the ASR of the round $t$ and $\lfloor\cdot\rfloor$ rounds down the float number.\\
\textit{\textbf{Utility degradation.}} Similarly, byzantine attacks~\cite{cao2022mpaf,LiNV23Byzantine,XuHSL22Byzantine} intend to disrupt the training process, exhibiting instability regarding the utility degradation (\textit{UD}, $ud_t$). Note that $ud_t$ is formulated as $acc_t-\hat{acc}_t$, where $acc_t$ and $\hat{acc}_t$ stand for the precision of the normally trained and poisoned global model in round $t$. Therefore, we used byzantine degradation accuracy (\textit{BDA}, $\mu_{ud}$) and variance (\textit{BDV}, $\sigma_{ud}$) of the last 10\% rounds to replace traditional UD metrics. \\

\textit{\textbf{Poison stealthiness.}} To achieve the above effectiveness while avoiding potential detection, previous attacks tend to validate the poison stealthiness via existing defenses~\cite{}, most of which fail to satisfy practical needs as stated in Sec~\ref{sec:fl_in_practice}. Therefore, for backdoor attacks, we calculate the main task accuracy and variance (ACCV) of the last 10\% rounds to ensure the training has converged to measure how the poisoning affects the main task.

\begin{takeaway}
\textbf{Gap III (Security).} Traditional metrics often fail to account for attack stability, utility degradation, and attack stealthiness in practical deployments. Similarly, utility degradation and poison stealthiness are often assessed using outdated metrics, which do not reflect the realistic dynamics of FL systems.
\end{takeaway}
\vspace{5pt}

% When presenting this system design to reviewers, consider the following key points:

% Clarity: Ensure that each component and its interaction within the framework are described clearly and coherently.
% Justification: Provide strong reasons for the design choices, especially how they simulate realistic FL environments and threats.
% Replicability: Ensure the system design is detailed enough to be replicated by other researchers, enhancing the reliability and reproducibility of our findings.
% Evaluation: Highlight how our performance metrics offer a practical assessment of FL poisoning-security risks.
% Significance: Emphasize the importance of bridging the gap between research methodologies and practical implementations, and how our framework contributes to this endeavor.
% By addressing these points, we can effectively communicate the robustness and practicality of your system design to the reviewers, ensuring that our empirical evaluation is seen as thorough, credible, and significant.

\begin{figure}
    \centering
    \includegraphics[width=\linewidth]{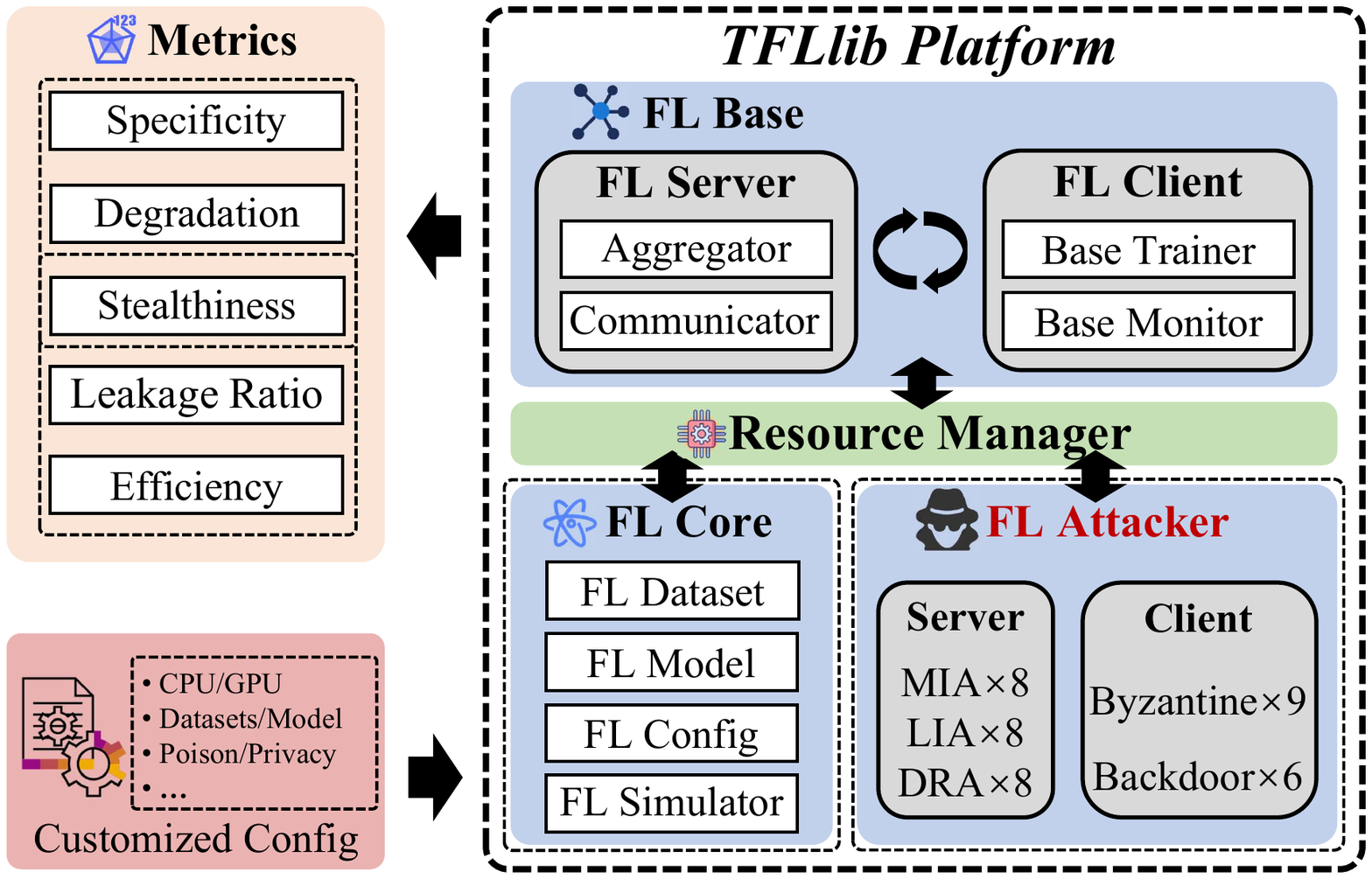}
    \caption{System Design of \textit{\textbf{TFLlib}}}
    \label{fig:system_design}
\end{figure}

\section{System Design and Implementation}
\label{sec:system}
To answer the \textbf{RQs} proposed before, we implement a measurement system for FL poisoning-security risks, \textit{\textbf{TFLlib}}, as shown in Fig~\ref{fig:system_design}. Our goal is to empirically validate the risk gaps identified earlier and derive actionable insights about how poisoning threats behave in contemporary FL frameworks under practical constraints.
\subsection{System Design}
Since existing platforms are poor at reproducing practical and security-relevant FL scenarios, we introduce \textit{\textbf{TFLlib}}, composed of three parts: \texttt{FL Base}, \texttt{FL Core}, and \texttt{FL Attacker}, together with the proposed metrics. Specifically, the \texttt{FL Base} module supports the minimal simulation of FL tasks across three modalities (i.e., tabular, image, and text) on three model families (i.e., MLPs, CNNs, and Transformers). Additionally, the \texttt{FL Attacker} inherits the parent classes \texttt{FL Client} to launch poisoning attacks.

To the best of our knowledge, \textit{\textbf{TFLlib}} is the first implemented uniform measurement system in this line of work that systematically integrates representative poisoning attacks and their utility-aware security metrics under one practical FL framework. \textit{\textbf{TFLlib}} offers FL researchers and practitioners a uniform simulation platform to bridge the gap between real-world scenarios and research configurations. Researchers can incorporate customized attacks to measure their effectiveness and stealthiness. Additionally, \textit{\textbf{TFLlib}} refines several impractical experimental settings that can otherwise lead to contradictory conclusions. In this way, \textit{\textbf{TFLlib}} reduces evaluation bias and facilitates fairer comparisons among attacks in the wild.

\subsection{System Implementation}

Our system framework, depicted in Fig~\ref{fig:system_design}, integrates the components needed for a practical evaluation of FL poisoning-security risks. The framework consists of:

\textbf{FL Benchmarks}. \textit{\textbf{TFLlib}} offers comprehensive datasets and models for evaluation. For simplicity but without loss of generality, we adopt CIFAR10~\cite{krizhevsky2009learning}, FEMNIST~\cite{leaf}, Purchase100~\cite{ShokriSSS17Membership}, Texas100~\cite{ShokriSSS17Membership}, IMDB~\cite{imdb} and AGNews~\cite{ZhangZL15} in this paper, covering image, tabular and text data.

\textbf{Hybrid Heterogeneity}. As mentioned before, clients exhibit data, resources, and communication heterogeneity simultaneously. Specifically, the local data for each client is sampled from a Dirichlet distribution \textit{Dir}($\tau^{s}$) and then split into training and testing parts, respectively. The test data from all the clients are aggregated together to evaluate the performance of the global model. Considering the practical distribution of devices~\cite{aibenchmark} and communication~\cite{YangWXCBLL21Characterizing,fedscale}, we use truncated Gaussian distribution to simulate real-world scenarios. Formally, we sample $\tau^{d}_{i}\sim\mathcal{N}_{[0,1]}(0,1-\tau^{d})$ and $\tau^{c}_{i}\sim\mathcal{N}_{[0,1]}(0,1-\tau^{c})$. The client with higher $\tau^{d}_{i}$ and $\tau^{c}_{i}$ will be assigned with less computation (lower $E_i$) and a higher possibility of rejection by aggregation.

\jiahao{Why we selected these attacks}
\textbf{Poisoning Attack}. This component exploits vulnerabilities in FL scenarios above with: six backdoor attacks including DBA~\cite{xie2019dba}, CerP~\cite{LyuHWLWL023Poisoning}, EdgeCase~\cite{WangSRVASLP20Attack}, A3FL~\cite{ZhangJCLW23A3FL}, Replace~\cite{BagdasaryanVHES20How} and Neurotoxin~\cite{ZhangPSYMMR022Neurotoxin}; nine byzantine attacks including IPM~\cite{XieKG19}, Noise~\cite{BlanchardMGS17,LiNV23Byzantine}, Fang~\cite{FangCJG20Local}, LabelFlip~\cite{FangCJG20Local}, SignGuard~\cite{XuHSL22Byzantine}, UpdateFlip, MinMax~\cite{ShejwalkarH21Manipulating}, MedianTailored, SignFlip~\cite{DamaskinosMGPT18,LiXCGL19RSA} and LIE~\cite{BaruchBG19A}. The detailed attack parameters are given in Appendix~\ref{sec:appendix_evaluated_attacks}. Notably, we refine the existing attack settings regarding the identified gaps in Sec~\ref{sec:gap1}. For instance, we provide malicious clients with updates from other malicious clients instead of benign clients. 

\textbf{Practical Evaluation}. For security evaluation under the identified gaps, we reformulate and construct objective-driven metrics from the adversary's perspective. These metrics highlight not only attack effectiveness but also stability and stealthiness under realistic FL conditions. Moreover, \textit{\textbf{TFLlib}} adopts a modular implementation with Pytorch~\cite{} and provides parallel modules for GPU acceleration, which makes it easily extendable and manageable.

\begin{figure*}
\centering
\includegraphics[width=0.99\linewidth]{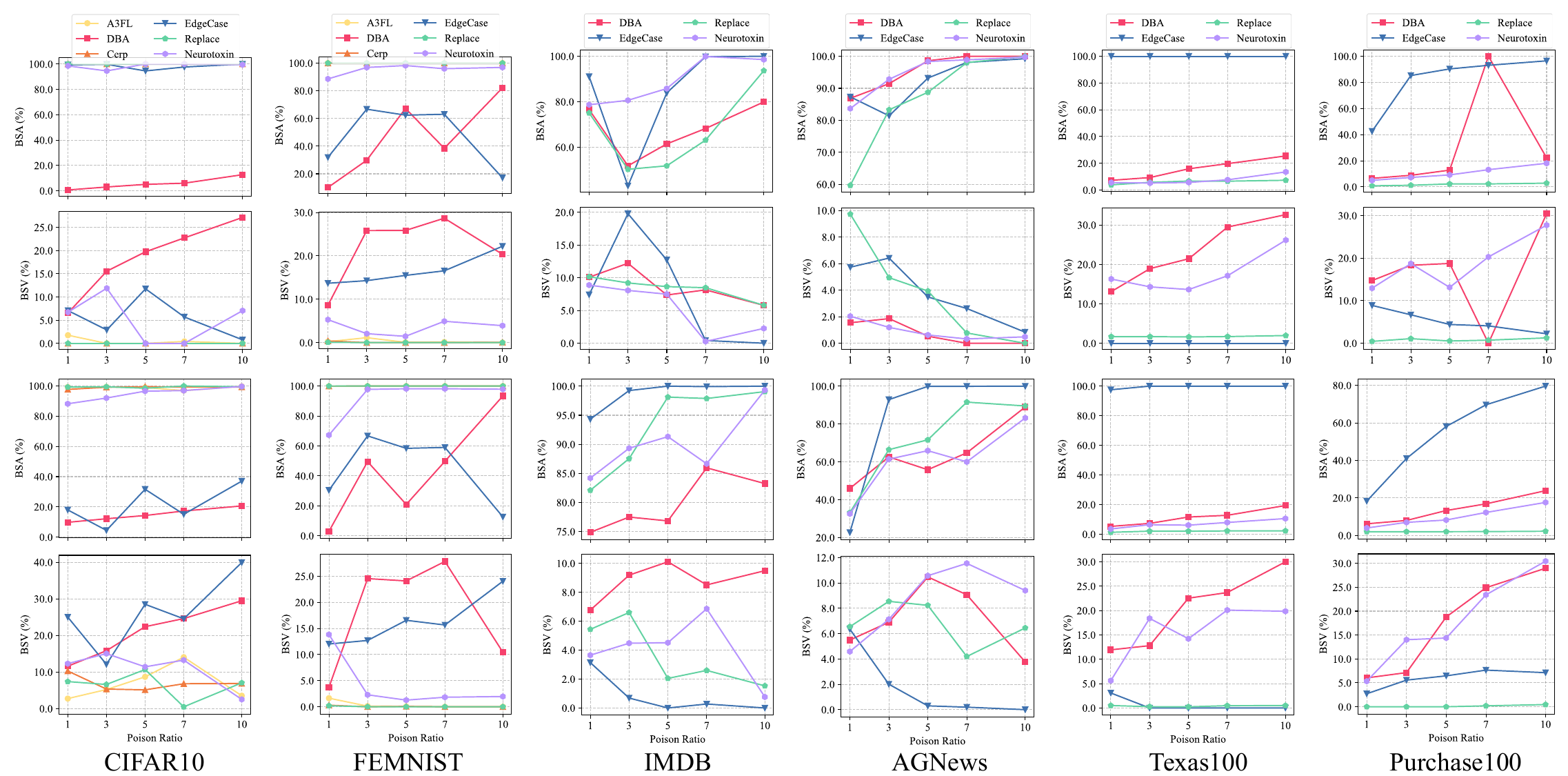}
\caption{Backdoor attack performance under practical (top two rows) and ideal (bottom two rows) settings. For each dataset, we report backdoor specificity accuracy (BSA) and backdoor specificity variance (BSV) under different poison ratios.}
\label{fig:main_bkd_asr}
\end{figure*}

\section{Evaluation}
\label{sec:evaluation}
In this section, we benchmark the \emph{security} risks of FL under practical scenarios with \textit{\textbf{TFLlib}} to answer the research questions above. For this arXiv version, we intentionally focus on poisoning-based security risks and omit the incomplete privacy evaluation from the main text to avoid over-claiming. We first quantify how much idealized settings overestimate poisoning attacks in deployed FL scenarios. We then explain why practical threat-model constraints and hybrid heterogeneity change the observed attack behavior. Finally, we show why practical security measurement must jointly consider effectiveness, stability, and collateral damage instead of reporting only a single end-of-training attack score.
% 对于posion attack, 我考虑两种threat model, 设定10,5,1%随机选择。
% 对于privacy attack, 对于reconstruction，考虑batch_size和数据集难度；对于label_inference，batch_size, multiple_local_epochs;对于MIA，辅助数据集的数量，验证集，target_client参与轮数。
% client:单个batch单轮，不同batchsize; 固定数据量和batchsize，不同local_epochs；辅助数据集数量，辅助数据集是否与原数据集同分布。
% 我们需要要在实验中寻找一些new threat, 可以尝试隐私和poison一起进行实验。比如由于数据分布或者模态带来的威胁，隐私可能更多一些
% 找一些例子，现有评估指标的不足，无法评估安全性。比如ASR,但是计算开销很大, 可能需要找出我们的实验结果与论文中的结果进行对比，区别在哪里，是否overclaim 
% 构建多个威胁模型，不同程度的knowledge和capability。
% 对于poison,考虑两种不同的threat model之间的区别
% 讨论异构参数，数据集对安全隐私的影响
% 讨论后门or拜占庭对隐私的影响
\subsection{Setup}
The datasets and models involved in the evaluation are given in Tab~\ref{tab:exp_cfg} and Appendix~\ref{sec:appendix_dataset_model}, with various global rounds $T$ and learning rates $\eta$. Specifically, we split 10\% of the whole datasets for FL global evaluation. Other FL parameters are given as follows: the number of the clients $N=100$; the percentage of the aggregated clients $\alpha=0.1$; local data batch size $B=64$ and local training epoch $E=5$. We also define hybrid heterogeneity of ``practical settings'' as $\tau^{s}=0.9, \tau^{c}=0.9, \tau^{d}=0.9$ and ``ideal settings'' as $\tau^{s}=\infty, \tau^{c}=1, \tau^{d}=1$ (except for FEMNIST).

\subsection{Poisoning Attacks}
In this subsection, we evaluate poisoning attacks under the practical threat model described in Sec~\ref{sec:gap1} and compare them against the commonly adopted idealized setting. Our goal is not merely to ask whether an attack can ever succeed, but whether it remains effective, stable, and sufficiently stealthy once realistic participation constraints and hybrid heterogeneity are introduced.

\textbf{Backdoor Attacks.}
\emph{Result.} Fig.~\ref{fig:main_bkd_asr} reveals a clear gap between idealized and practical evaluation. Under ideal settings, several attacks appear almost universally strong: their BSA is close to saturation on many datasets and their behavior is comparatively smooth across poison ratios. This picture changes substantially under practical settings. The achieved BSA becomes highly attack- and dataset-dependent, and the gap is especially visible on the tabular benchmarks. On Texas100, for example, EdgeCase remains consistently effective, whereas DBA, Replace, and Neurotoxin stay at much lower BSA across all poison ratios. On Purchase100, DBA can momentarily reach a high BSA, but the trend is strongly non-monotonic, while Replace remains largely ineffective. In contrast, on image and text benchmarks such as CIFAR10, IMDB, and AGNews, several attacks still achieve high BSA, showing that practical constraints do not eliminate the backdoor risk, but they do make it much less uniform than prior idealized evaluations suggest.

\emph{Explanation.} The discrepancy follows directly from the gaps identified in Sec~\ref{sec:motivation}. In practical FL, attackers do not participate in every round, cannot rely on a large compromised population, and must survive statistical, device, and communication heterogeneity. These factors weaken coordinated poisoning and make the effect of poisoned updates less persistent. As a result, attacks that look broadly successful in IID and fully controlled simulations can become brittle once poisoned clients are randomly sampled and trained under heterogeneous conditions. The practical curves also show frequent non-monotonicity with respect to the poison ratio, indicating that simply increasing the attacker budget does not guarantee a stronger or more reliable attack in realistic FL deployments.

\emph{Result.} Practical evaluation also changes the stability story. The second and fourth rows of Fig.~\ref{fig:main_bkd_asr} show that attacks with a competitive BSA can still exhibit a large BSV, especially on CIFAR10, FEMNIST, Texas100, and Purchase100. This means that a high final-round ASR may be produced by a narrow and unstable operating region instead of a durable attack. Appendix Fig.~\ref{fig:main_bkd_acc} further shows that high attack specificity can come with visible degradation of the main task or large ACCV. A representative example is DBA on Purchase100: although it reaches a high BSA at one practical poison ratio, benign accuracy collapses sharply at the same point, making the attack difficult to hide in a deployed system. By contrast, the more concerning attacks in practice are those that maintain both a high BSA and a comparatively small utility penalty, such as EdgeCase on Texas100 and the stronger attacks on IMDB/AGNews.

\emph{Explanation.} These observations expose why end-of-training ASR alone is an incomplete measurement. Practical attackers care about \emph{durable} and \emph{stealthy} influence, not just a single successful round. Once benign utility and temporal stability are included, the ranking between attacks can change substantially. In other words, idealized evaluation overstates not only the magnitude of the threat, but also which attacks are actually practical.

\textbf{Scope of this arXiv version.} The backdoor benchmark is currently the most complete empirical component of \textit{\textbf{TFLlib}}, so we use it as the representative poisoning case study in the main paper. The same practical constraints also apply to byzantine attacks, and the framework already measures them with BDA/BDV-style metrics, but we omit incomplete per-attack byzantine plots from this version to keep the main text evidence-aligned.

\begin{takeaway} 
\textbf{RQ1 (Security).} Existing FL security evaluations substantially overestimate poisoning risk when they assume idealized participation, homogeneous clients, and single-number success metrics. Under practical conditions, backdoor attacks remain a real threat, but their effectiveness is far more selective, unstable, and coupled with utility loss than the literature often implies.
\end{takeaway}

\subsection{Security Measurement of FL Frameworks}
\emph{Result.} Our practical measurement highlights that attack effectiveness, stability, and stealthiness must be reported together. Fig.~\ref{fig:main_bkd_asr} and Appendix Fig.~\ref{fig:main_bkd_acc} show several representative failure modes of single-metric reporting. First, peak or final-round ASR can exaggerate attacks whose performance spikes only in a narrow range of poison ratios. Second, a strong BSA may coexist with a high BSV, meaning the attack is difficult to reproduce reliably across the last stage of training. Third, a seemingly successful attack can impose such a severe benign-accuracy penalty that it becomes operationally obvious. These cases would all be mischaracterized if the evaluation only reported the final ASR.

\emph{Explanation.} This is precisely why we reformulate practical security measurement around objective-driven metrics. For backdoor attacks, BSA captures sustained specificity in the converged phase, BSV measures whether that specificity is stable, and ACC/ACCV expose the cost paid on the benign task. For byzantine attacks, the same logic motivates BDA/BDV instead of single-round degradation. These metrics prevent cherry-picking transient peaks and make cross-attack comparisons fairer under heterogeneous FL settings.

\begin{takeaway}
\textbf{Takeaway (Measurement).} Practical FL security should be reported as a joint tuple of \emph{effectiveness + stability + utility cost}. Metrics based only on the final-round success rate systematically overstate practical risk and can even reverse the ranking between attacks once temporal variance and benign-task degradation are taken into account.
\end{takeaway}

% \section{Discussion}
% \label{sec:discussion}

\section{Future Work}
\label{sec:future_work}

This arXiv version focuses on poisoning-based security risks and leaves privacy-specific FL risks for future work. In particular, an important next step is to extend \textit{\textbf{TFLlib}} to server-side privacy attacks, such as inversion and inference attacks, and to evaluate how practical deployment choices affect the true severity of those threats. The same research-to-practice gap may exist there as well: many privacy-preserving techniques proposed in the literature are still rarely integrated into production FL stacks, and their real utility-cost tradeoff under heterogeneous deployments remains unclear.

Another important direction is to study how privacy risks interact with model and modality diversity. Real FL workloads span MLPs, CNNs, and Transformers across tabular, image, and text tasks, and privacy behavior may vary substantially across these settings. A future privacy benchmark should therefore compare attack efficacy and defense cost under the same practical principles adopted in this paper: realistic participation, heterogeneous clients, and utility-aware measurement.

\section{Implications}
\label{sec:implications}

The results above suggest that the central problem is not whether poisoning attacks exist, but how they should be assessed and acted upon under practical FL constraints. Our findings therefore have direct implications for two groups: researchers who design attacks and defenses, and practitioners who deploy FL systems in the wild.

\subsection{For FL Algorithm Designers}

The first implication for FL algorithm designers is that practical threat modeling should be treated as a first-class design constraint rather than a post-hoc evaluation choice. Attack papers should no longer assume fixed attacker participation, unrealistically high malicious ratios, or homogeneous clients by default, because these assumptions can substantially overstate the attack surface. Likewise, defense papers should not claim practical robustness if they are only validated against attacks that are artificially strengthened by idealized participation or simplified system settings.

The second implication is that security evaluation must move beyond a single end-of-training success number. Our results show that practical poisoning behavior depends on dataset modality, model family, client heterogeneity, and the temporal stability of the attack. Consequently, future FL security work should report effectiveness, stability, and benign-task degradation together, and should compare methods under a common set of realistic configurations. Otherwise, the community risks optimizing for attacks that are visually strong in controlled experiments but brittle in deployment, or for defenses that appear effective only because the benchmark itself is misaligned with practice.

Finally, our study suggests that future attack and defense design should target robustness to heterogeneity rather than performance in a single narrow regime. In realistic FL systems, statistical skew, device variability, and communication failures are not noise around the experiment; they are part of the attack surface. Algorithms that remain effective or robust under these conditions are substantially more meaningful than those tuned to IID, synchronous, and fully reliable settings.

\subsection{For FL System Providers}

For FL system providers, the main implication is that academic attack numbers should not be interpreted as deployment risk without checking the underlying assumptions. In particular, attack success reported under fixed attacker participation or homogeneous clients can exaggerate the real threat faced by a production FL service. System providers should therefore evaluate poisoning risk using their own client population, participation policy, modality mix, and benign-performance constraints before translating research claims into operational decisions.

Our measurements also suggest that system providers should monitor more than whether a backdoor can be triggered at the end of training. Practical attacks that appear strong on a single metric may still be unstable or operationally visible because they degrade benign accuracy or create abnormal variance over time. This means that deployment monitoring should track not only task accuracy, but also temporal instability and suspicious utility loss, since these signals can reveal attacks that would look successful in a paper but are difficult to sustain quietly in production.

More broadly, the results support a conservative and measurement-driven deployment strategy. Instead of adopting heavy defenses solely because they are state-of-the-art in the literature, providers should prioritize evaluation protocols that reflect their actual system conditions and identify where the largest practical risk really lies. In many cases, realistic client sampling, heterogeneity-aware testing, and utility-aware monitoring may offer more actionable security value than relying on conclusions drawn from idealized benchmarks alone.

Taken together, these implications point to the same message: practical FL security is a measurement problem before it is an optimization problem. If researchers and practitioners align on realistic assumptions, heterogeneous settings, and utility-aware metrics, the field can move from exaggerated attack narratives toward a more credible understanding of which poisoning threats actually matter in deployed federated learning.

{
\footnotesize 
\bibliographystyle{acm}
\bibliography{usenix}

\begin{thebibliography}{10}

\bibitem{ApruzzeseADFPR23Real}
{\sc Apruzzese, G., Anderson, H.~S., Dambra, S., Freeman, D., Pierazzi, F., and Roundy, K.~A.}
\newblock "real attackers don't compute gradients": Bridging the gap between adversarial {ML} research and practice.
\newblock In {\em 2023 {IEEE} Conference on Secure and Trustworthy Machine Learning, SaTML 2023, Raleigh, NC, USA, February 8-10, 2023\/} (2023), {IEEE}, pp.~339--364.

\bibitem{BagdasaryanVHES20How}
{\sc Bagdasaryan, E., Veit, A., Hua, Y., Estrin, D., and Shmatikov, V.}
\newblock How to backdoor federated learning.
\newblock In {\em The 23rd International Conference on Artificial Intelligence and Statistics, {AISTATS} 2020, 26-28 August 2020, Online [Palermo, Sicily, Italy]\/} (2020), S.~Chiappa and R.~Calandra, Eds., vol.~108 of {\em Proceedings of Machine Learning Research}, {PMLR}, pp.~2938--2948.

\bibitem{paddlefl}
{\sc BaiDu}.
\newblock Federated deep learning in paddlepaddle.
\newblock \url{https://github.com/PaddlePaddle/PaddleFL}, 2017.

\bibitem{BaruchBG19A}
{\sc Baruch, G., Baruch, M., and Goldberg, Y.}
\newblock A little is enough: Circumventing defenses for distributed learning.
\newblock In {\em Advances in Neural Information Processing Systems 32: Annual Conference on Neural Information Processing Systems 2019, NeurIPS 2019, December 8-14, 2019, Vancouver, BC, Canada\/} (2019), H.~M. Wallach, H.~Larochelle, A.~Beygelzimer, F.~d'Alch{\'{e}}{-}Buc, E.~B. Fox, and R.~Garnett, Eds., pp.~8632--8642.

\bibitem{flower}
{\sc Beutel, D.~J., Topal, T., Mathur, A., Qiu, X., Parcollet, T., and Lane, N.~D.}
\newblock Flower: {A} friendly federated learning research framework.
\newblock {\em CoRR abs/2007.14390\/} (2020).

\bibitem{BlanchardMGS17Machine}
{\sc Blanchard, P., Mhamdi, E. M.~E., Guerraoui, R., and Stainer, J.}
\newblock Machine learning with adversaries: Byzantine tolerant gradient descent.
\newblock In {\em Advances in Neural Information Processing Systems 30: Annual Conference on Neural Information Processing Systems 2017, December 4-9, 2017, Long Beach, CA, {USA}\/} (2017), I.~Guyon, U.~von Luxburg, S.~Bengio, H.~M. Wallach, R.~Fergus, S.~V.~N. Vishwanathan, and R.~Garnett, Eds., pp.~119--129.

\bibitem{BlanchardMGS17}
{\sc Blanchard, P., Mhamdi, E. M.~E., Guerraoui, R., and Stainer, J.}
\newblock Machine learning with adversaries: Byzantine tolerant gradient descent.
\newblock In {\em Advances in Neural Information Processing Systems 30: Annual Conference on Neural Information Processing Systems 2017, December 4-9, 2017, Long Beach, CA, {USA}\/} (2017), I.~Guyon, U.~von Luxburg, S.~Bengio, H.~M. Wallach, R.~Fergus, S.~V.~N. Vishwanathan, and R.~Garnett, Eds., pp.~119--129.

\bibitem{bonawitz2019towards}
{\sc Bonawitz, K., Eichner, H., Grieskamp, W., Huba, D., Ingerman, A., Ivanov, V., Kiddon, C., Kone{\v{c}}n{\`y}, J., Mazzocchi, S., McMahan, B., et~al.}
\newblock Towards federated learning at scale: System design.
\newblock {\em Proceedings of machine learning and systems 1\/} (2019), 374--388.

\bibitem{bottou2010large}
{\sc Bottou, L.}
\newblock Large-scale machine learning with stochastic gradient descent.
\newblock In {\em Proceedings of COMPSTAT'2010: 19th International Conference on Computational StatisticsParis France, August 22-27, 2010 Keynote, Invited and Contributed Papers\/} (2010), Springer, pp.~177--186.

\bibitem{fedlearner}
{\sc ByteDance}.
\newblock A multi-party collaborative machine learning framework.
\newblock \url{https://github.com/bytedance/fedlearner}, 2020.

\bibitem{leaf}
{\sc Caldas, S., Duddu, S. M.~K., Wu, P., Li, T., Kone{\v{c}}n{\`y}, J., McMahan, H.~B., Smith, V., and Talwalkar, A.}
\newblock Leaf: A benchmark for federated settings.
\newblock {\em arXiv preprint arXiv:1812.01097\/} (2018).

\bibitem{CaoF0G21FLTrust}
{\sc Cao, X., Fang, M., Liu, J., and Gong, N.~Z.}
\newblock Fltrust: Byzantine-robust federated learning via trust bootstrapping.
\newblock In {\em 28th Annual Network and Distributed System Security Symposium, {NDSS} 2021, virtually, February 21-25, 2021\/} (2021), The Internet Society.

\bibitem{cao2022mpaf}
{\sc Cao, X., and Gong, N.~Z.}
\newblock Mpaf: Model poisoning attacks to federated learning based on fake clients.
\newblock In {\em Proceedings of the IEEE/CVF Conference on Computer Vision and Pattern Recognition\/} (2022), pp.~3396--3404.

\bibitem{cao2023fedrecover}
{\sc Cao, X., Jia, J., Zhang, Z., and Gong, N.~Z.}
\newblock Fedrecover: Recovering from poisoning attacks in federated learning using historical information.
\newblock In {\em 2023 IEEE Symposium on Security and Privacy (SP)\/} (2023), IEEE, pp.~1366--1383.

\bibitem{DamaskinosMGPT18}
{\sc Damaskinos, G., Mhamdi, E. M.~E., Guerraoui, R., Patra, R., and Taziki, M.}
\newblock Asynchronous byzantine machine learning (the case of {SGD)}.
\newblock In {\em Proceedings of the 35th International Conference on Machine Learning, {ICML} 2018, Stockholmsm{\"{a}}ssan, Stockholm, Sweden, July 10-15, 2018\/} (2018), J.~G. Dy and A.~Krause, Eds., vol.~80 of {\em Proceedings of Machine Learning Research}, {PMLR}, pp.~1153--1162.

\bibitem{aibenchmark}
{\sc ETHZurich}.
\newblock Ai benchmark.
\newblock \url{https://ai-benchmark.com/ranking.html}, 2019.

\bibitem{FangCJG20Local}
{\sc Fang, M., Cao, X., Jia, J., and Gong, N.~Z.}
\newblock Local model poisoning attacks to byzantine-robust federated learning.
\newblock In {\em 29th {USENIX} Security Symposium, {USENIX} Security 2020, August 12-14, 2020\/} (2020), S.~Capkun and F.~Roesner, Eds., {USENIX} Association, pp.~1605--1622.

\bibitem{garcia2022flute}
{\sc Garcia, M.~H., Manoel, A., Diaz, D.~M., Mireshghallah, F., Sim, R., and Dimitriadis, D.}
\newblock Flute: A scalable, extensible framework for high-performance federated learning simulations.
\newblock {\em arXiv preprint arXiv:2203.13789\/} (2022).

\bibitem{tff}
{\sc Google}.
\newblock An open-source framework for machine learning and other computations on decentralized data.
\newblock \url{https://github.com/google-parfait/tensorflow-federated}, 2017.

\bibitem{fedml}
{\sc He, C., Li, S., So, J., Zhang, M., Wang, H., Wang, X., Vepakomma, P., Singh, A., Qiu, H., Shen, L., Zhao, P., Kang, Y., Liu, Y., Raskar, R., Yang, Q., Annavaram, M., and Avestimehr, S.}
\newblock Fedml: {A} research library and benchmark for federated machine learning.
\newblock {\em CoRR abs/2007.13518\/} (2020).

\bibitem{HeZRS16}
{\sc He, K., Zhang, X., Ren, S., and Sun, J.}
\newblock Deep residual learning for image recognition.
\newblock In {\em 2016 {IEEE} Conference on Computer Vision and Pattern Recognition, {CVPR} 2016, Las Vegas, NV, USA, June 27-30, 2016\/} (2016), {IEEE} Computer Society, pp.~770--778.

\bibitem{mindsporefl}
{\sc Huawei}.
\newblock Mindspore federated.
\newblock \url{https://gitee.com/mindspore/federated}, 2022.

\bibitem{huba2022papaya}
{\sc Huba, D., Nguyen, J., Malik, K., Zhu, R., Rabbat, M., Yousefpour, A., Wu, C.-J., Zhan, H., Ustinov, P., Srinivas, H., et~al.}
\newblock Papaya: Practical, private, and scalable federated learning.
\newblock {\em Proceedings of Machine Learning and Systems 4\/} (2022), 814--832.

\bibitem{imdb}
{\sc IMDbPro}.
\newblock Imdb data files available for download.
\newblock \url{https://datasets.imdbws.com/}, 2022.

\bibitem{ImteajTWLA22A}
{\sc Imteaj, A., Thakker, U., Wang, S., Li, J., and Amini, M.~H.}
\newblock A survey on federated learning for resource-constrained iot devices.
\newblock {\em {IEEE} Internet Things J. 9}, 1 (2022), 1--24.

\bibitem{JiaoYSJCL0L20}
{\sc Jiao, X., Yin, Y., Shang, L., Jiang, X., Chen, X., Li, L., Wang, F., and Liu, Q.}
\newblock Tinybert: Distilling {BERT} for natural language understanding.
\newblock In {\em Findings of the Association for Computational Linguistics: {EMNLP} 2020, Online Event, 16-20 November 2020\/} (2020), T.~Cohn, Y.~He, and Y.~Liu, Eds., vol.~{EMNLP} 2020 of {\em Findings of {ACL}}, Association for Computational Linguistics, pp.~4163--4174.

\bibitem{KaissisMRB20Secure}
{\sc Kaissis, G., Makowski, M.~R., Rueckert, D., and Braren, R.}
\newblock Secure, privacy-preserving and federated machine learning in medical imaging.
\newblock {\em Nat. Mach. Intell. 2}, 6 (2020), 305--311.

\bibitem{KarimireddyHJ21Learning}
{\sc Karimireddy, S.~P., He, L., and Jaggi, M.}
\newblock Learning from history for byzantine robust optimization.
\newblock In {\em Proceedings of the 38th International Conference on Machine Learning, {ICML} 2021, 18-24 July 2021, Virtual Event\/} (2021), M.~Meila and T.~Zhang, Eds., vol.~139 of {\em Proceedings of Machine Learning Research}, {PMLR}, pp.~5311--5319.

\bibitem{khan2023pitfalls}
{\sc Khan, M.~A., Shejwalkar, V., Houmansadr, A., and Anwar, F.~M.}
\newblock On the pitfalls of security evaluation of robust federated learning.
\newblock In {\em 2023 IEEE Security and Privacy Workshops (SPW)\/} (2023), IEEE, pp.~57--68.

\bibitem{KhanSHA23On}
{\sc Khan, M.~A., Shejwalkar, V., Houmansadr, A., and Anwar, F.~M.}
\newblock On the pitfalls of security evaluation of robust federated learning.
\newblock In {\em 2023 {IEEE} Security and Privacy Workshops (SPW), San Francisco, CA, USA, May 25, 2023\/} (2023), {IEEE}, pp.~57--68.

\bibitem{krizhevsky2009learning}
{\sc Krizhevsky, A., Hinton, G., et~al.}
\newblock Learning multiple layers of features from tiny images.

\bibitem{fedscale}
{\sc Lai, F., Dai, Y., Singapuram, S. S.~V., Liu, J., Zhu, X., Madhyastha, H.~V., and Chowdhury, M.}
\newblock Fedscale: Benchmarking model and system performance of federated learning at scale.
\newblock In {\em International Conference on Machine Learning, {ICML} 2022, 17-23 July 2022, Baltimore, Maryland, {USA}\/} (2022), K.~Chaudhuri, S.~Jegelka, L.~Song, C.~Szepesv{\'{a}}ri, G.~Niu, and S.~Sabato, Eds., vol.~162 of {\em Proceedings of Machine Learning Research}, {PMLR}, pp.~11814--11827.

\bibitem{LiYHLWFS233DFed}
{\sc Li, H., Ye, Q., Hu, H., Li, J., Wang, L., Fang, C., and Shi, J.}
\newblock 3dfed: Adaptive and extensible framework for covert backdoor attack in federated learning.
\newblock In {\em 44th {IEEE} Symposium on Security and Privacy, {SP} 2023, San Francisco, CA, USA, May 21-25, 2023\/} (2023), {IEEE}, pp.~1893--1907.

\bibitem{LiXCGL19RSA}
{\sc Li, L., Xu, W., Chen, T., Giannakis, G.~B., and Ling, Q.}
\newblock {RSA:} byzantine-robust stochastic aggregation methods for distributed learning from heterogeneous datasets.
\newblock In {\em The Thirty-Third {AAAI} Conference on Artificial Intelligence, {AAAI} 2019, The Thirty-First Innovative Applications of Artificial Intelligence Conference, {IAAI} 2019, The Ninth {AAAI} Symposium on Educational Advances in Artificial Intelligence, {EAAI} 2019, Honolulu, Hawaii, USA, January 27 - February 1, 2019\/} (2019), {AAAI} Press, pp.~1544--1551.

\bibitem{LiNV23Byzantine}
{\sc Li, S., Ngai, E. C.~H., and Voigt, T.}
\newblock Byzantine-robust aggregation in federated learning empowered industrial iot.
\newblock {\em {IEEE} Trans. Ind. Informatics 19}, 2 (2023), 1165--1175.

\bibitem{LiSBS20Fair}
{\sc Li, T., Sanjabi, M., Beirami, A., and Smith, V.}
\newblock Fair resource allocation in federated learning.
\newblock In {\em 8th International Conference on Learning Representations, {ICLR} 2020, Addis Ababa, Ethiopia, April 26-30, 2020\/} (2020), OpenReview.net.

\bibitem{unifed}
{\sc Liu, X., Shi, T., Xie, C., Li, Q., Hu, K., Kim, H., Xu, X., Li, B., and Song, D.}
\newblock Unifed: A benchmark for federated learning frameworks.
\newblock {\em arXiv preprint arXiv:2207.10308\/} (2022).

\bibitem{fate}
{\sc Liu, Y., Fan, T., Chen, T., Xu, Q., and Yang, Q.}
\newblock {FATE:} an industrial grade platform for collaborative learning with data protection.
\newblock {\em J. Mach. Learn. Res. 22\/} (2021), 226:1--226:6.

\bibitem{LongT0Z20Federated}
{\sc Long, G., Tan, Y., Jiang, J., and Zhang, C.}
\newblock Federated learning for open banking.
\newblock In {\em Federated Learning - Privacy and Incentive}, Q.~Yang, L.~Fan, and H.~Yu, Eds., vol.~12500 of {\em Lecture Notes in Computer Science}. Springer, 2020, pp.~240--254.

\bibitem{ibmfl}
{\sc Ludwig, H., Baracaldo, N., Thomas, G., Zhou, Y., Anwar, A., Rajamoni, S., Ong, Y.~J., Radhakrishnan, J., Verma, A., Sinn, M., Purcell, M., Rawat, A., Minh, T.~N., Holohan, N., Chakraborty, S., Witherspoon, S., Steuer, D., Wynter, L., Hassan, H., Laguna, S., Yurochkin, M., Agarwal, M., Chuba, E., and Abay, A.}
\newblock {IBM} federated learning: an enterprise framework white paper {V0.1}.
\newblock {\em CoRR abs/2007.10987\/} (2020).

\bibitem{LycklamaBVKH23RoFL}
{\sc Lycklama, H., Burkhalter, L., Viand, A., K{\"{u}}chler, N., and Hithnawi, A.}
\newblock Rofl: Robustness of secure federated learning.
\newblock In {\em 44th {IEEE} Symposium on Security and Privacy, {SP} 2023, San Francisco, CA, USA, May 21-25, 2023\/} (2023), {IEEE}, pp.~453--476.

\bibitem{LyuHWLWL023Poisoning}
{\sc Lyu, X., Han, Y., Wang, W., Liu, J., Wang, B., Liu, J., and Zhang, X.}
\newblock Poisoning with cerberus: Stealthy and colluded backdoor attack against federated learning.
\newblock In {\em Thirty-Seventh {AAAI} Conference on Artificial Intelligence, {AAAI} 2023, Thirty-Fifth Conference on Innovative Applications of Artificial Intelligence, {IAAI} 2023, Thirteenth Symposium on Educational Advances in Artificial Intelligence, {EAAI} 2023, Washington, DC, USA, February 7-14, 2023\/} (2023), B.~Williams, Y.~Chen, and J.~Neville, Eds., {AAAI} Press, pp.~9020--9028.

\bibitem{MansouriOJC23SoK}
{\sc Mansouri, M., {\"{O}}nen, M., Jaballah, W.~B., and Conti, M.}
\newblock Sok: Secure aggregation based on cryptographic schemes for federated learning.
\newblock {\em Proc. Priv. Enhancing Technol. 2023}, 1 (2023), 140--157.

\bibitem{mcmahan2017communication}
{\sc McMahan, B., Moore, E., Ramage, D., Hampson, S., and y~Arcas, B.~A.}
\newblock Communication-efficient learning of deep networks from decentralized data.
\newblock In {\em Artificial intelligence and statistics\/} (2017), PMLR, pp.~1273--1282.

\bibitem{MhamdiGR18The}
{\sc Mhamdi, E. M.~E., Guerraoui, R., and Rouault, S.}
\newblock The hidden vulnerability of distributed learning in byzantium.
\newblock In {\em Proceedings of the 35th International Conference on Machine Learning, {ICML} 2018, Stockholmsm{\"{a}}ssan, Stockholm, Sweden, July 10-15, 2018\/} (2018), J.~G. Dy and A.~Krause, Eds., vol.~80 of {\em Proceedings of Machine Learning Research}, {PMLR}, pp.~3518--3527.

\bibitem{MozaffariCH24Fake}
{\sc Mozaffari, H., Choudhary, S., and Houmansadr, A.}
\newblock Fake or compromised? making sense of malicious clients in federated learning.
\newblock In {\em Computer Security - {ESORICS} 2024 - 29th European Symposium on Research in Computer Security, Bydgoszcz, Poland, September 16-20, 2024, Proceedings, Part {I}\/} (2024), J.~Garc{\'{\i}}a{-}Alfaro, R.~Kozik, M.~Choras, and S.~K. Katsikas, Eds., vol.~14982 of {\em Lecture Notes in Computer Science}, Springer, pp.~187--207.

\bibitem{Luis19Byzantine}
{\sc Mu{\~{n}}oz{-}Gonz{\'{a}}lez, L., Co, K.~T., and Lupu, E.~C.}
\newblock Byzantine-robust federated machine learning through adaptive model averaging.
\newblock {\em CoRR abs/1909.05125\/} (2019).

\bibitem{NguyenDPSLP21Federated}
{\sc Nguyen, D.~C., Ding, M., Pathirana, P.~N., Seneviratne, A., Li, J., and Poor, H.~V.}
\newblock Federated learning for internet of things: {A} comprehensive survey.
\newblock {\em {IEEE} Commun. Surv. Tutorials 23}, 3 (2021), 1622--1658.

\bibitem{NguyenPPDSLDH23Federated}
{\sc Nguyen, D.~C., Pham, Q., Pathirana, P.~N., Ding, M., Seneviratne, A., Lin, Z., Dobre, O.~A., and Hwang, W.}
\newblock Federated learning for smart healthcare: {A} survey.
\newblock {\em {ACM} Comput. Surv. 55}, 3 (2023), 60:1--60:37.

\bibitem{ParkHCM21}
{\sc Park, J., Han, D., Choi, M., and Moon, J.}
\newblock Sageflow: Robust federated learning against both stragglers and adversaries.
\newblock In {\em Advances in Neural Information Processing Systems 34: Annual Conference on Neural Information Processing Systems 2021, NeurIPS 2021, December 6-14, 2021, virtual\/} (2021), M.~Ranzato, A.~Beygelzimer, Y.~N. Dauphin, P.~Liang, and J.~W. Vaughan, Eds., pp.~840--851.

\bibitem{QammarDN22Federated}
{\sc Qammar, A., Ding, J., and Ning, H.}
\newblock Federated learning attack surface: taxonomy, cyber defences, challenges, and future directions.
\newblock {\em Artif. Intell. Rev. 55}, 5 (2022), 3569--3606.

\bibitem{ReisizadehTHMP22}
{\sc Reisizadeh, A., Tziotis, I., Hassani, H., Mokhtari, A., and Pedarsani, R.}
\newblock Straggler-resilient federated learning: Leveraging the interplay between statistical accuracy and system heterogeneity.
\newblock {\em {IEEE} J. Sel. Areas Inf. Theory 3}, 2 (2022), 197--205.

\bibitem{pysyft}
{\sc Ryffel, T., Trask, A., Dahl, M., Wagner, B., Mancuso, J., Rueckert, D., and Passerat-Palmbach, J.}
\newblock A generic framework for privacy preserving deep learning.
\newblock {\em arXiv preprint arXiv:1811.04017\/} (2018).

\bibitem{SandeepaSWL24SHERPA}
{\sc Sandeepa, C., Siniarski, B., Wang, S., and Liyanage, M.}
\newblock {SHERPA:} explainable robust algorithms for privacy-preserved federated learning in future networks to defend against data poisoning attacks.
\newblock In {\em {IEEE} Symposium on Security and Privacy, {SP} 2024, San Francisco, CA, USA, May 19-23, 2024\/} (2024), {IEEE}, pp.~4772--4790.

\bibitem{ShejwalkarH21Manipulating}
{\sc Shejwalkar, V., and Houmansadr, A.}
\newblock Manipulating the byzantine: Optimizing model poisoning attacks and defenses for federated learning.
\newblock In {\em 28th Annual Network and Distributed System Security Symposium, {NDSS} 2021, virtually, February 21-25, 2021\/} (2021), The Internet Society.

\bibitem{shejwalkar2022back}
{\sc Shejwalkar, V., Houmansadr, A., Kairouz, P., and Ramage, D.}
\newblock Back to the drawing board: A critical evaluation of poisoning attacks on production federated learning.
\newblock In {\em 2022 IEEE Symposium on Security and Privacy (SP)\/} (2022), IEEE, pp.~1354--1371.

\bibitem{ShokriSSS17Membership}
{\sc Shokri, R., Stronati, M., Song, C., and Shmatikov, V.}
\newblock Membership inference attacks against machine learning models.
\newblock In {\em 2017 {IEEE} Symposium on Security and Privacy, {SP} 2017, San Jose, CA, USA, May 22-26, 2017\/} (2017), {IEEE} Computer Society, pp.~3--18.

\bibitem{WangSRVASLP20Attack}
{\sc Wang, H., Sreenivasan, K., Rajput, S., Vishwakarma, H., Agarwal, S., Sohn, J., Lee, K., and Papailiopoulos, D.~S.}
\newblock Attack of the tails: Yes, you really can backdoor federated learning.
\newblock In {\em Advances in Neural Information Processing Systems 33: Annual Conference on Neural Information Processing Systems 2020, NeurIPS 2020, December 6-12, 2020, virtual\/} (2020), H.~Larochelle, M.~Ranzato, R.~Hadsell, M.~Balcan, and H.~Lin, Eds.

\bibitem{xie2019dba}
{\sc Xie, C., Huang, K., Chen, P., and Li, B.}
\newblock {DBA:} distributed backdoor attacks against federated learning.
\newblock In {\em 8th International Conference on Learning Representations, {ICLR} 2020, Addis Ababa, Ethiopia, April 26-30, 2020\/} (2020), OpenReview.net.

\bibitem{XieKG19}
{\sc Xie, C., Koyejo, O., and Gupta, I.}
\newblock Fall of empires: Breaking byzantine-tolerant {SGD} by inner product manipulation.
\newblock In {\em Proceedings of the Thirty-Fifth Conference on Uncertainty in Artificial Intelligence, {UAI} 2019, Tel Aviv, Israel, July 22-25, 2019\/} (2019), A.~Globerson and R.~Silva, Eds., vol.~115 of {\em Proceedings of Machine Learning Research}, {AUAI} Press, pp.~261--270.

\bibitem{XieKG19Fall}
{\sc Xie, C., Koyejo, O., and Gupta, I.}
\newblock Fall of empires: Breaking byzantine-tolerant {SGD} by inner product manipulation.
\newblock In {\em Proceedings of the Thirty-Fifth Conference on Uncertainty in Artificial Intelligence, {UAI} 2019, Tel Aviv, Israel, July 22-25, 2019\/} (2019), A.~Globerson and R.~Silva, Eds., vol.~115 of {\em Proceedings of Machine Learning Research}, {AUAI} Press, pp.~261--270.

\bibitem{PoisonedFL}
{\sc Xie, Y., Fang, M., and Gong, N.~Z.}
\newblock Poisonedfl: Model poisoning attacks to federated learning via multi-round consistency.
\newblock {\em CoRR abs/2404.15611\/} (2024).

\bibitem{fedscope}
{\sc Xie, Y., Wang, Z., Chen, D., Gao, D., Yao, L., Kuang, W., Li, Y., Ding, B., and Zhou, J.}
\newblock Federatedscope: {A} comprehensive and flexible federated learning platform via message passing.
\newblock {\em CoRR abs/2204.05011\/} (2022).

\bibitem{XuHSL22Byzantine}
{\sc Xu, J., Huang, S., Song, L., and Lan, T.}
\newblock Byzantine-robust federated learning through collaborative malicious gradient filtering.
\newblock In {\em 42nd {IEEE} International Conference on Distributed Computing Systems, {ICDCS} 2022, Bologna, Italy, July 10-13, 2022\/} (2022), {IEEE}, pp.~1223--1235.

\bibitem{YangWXCBLL21Characterizing}
{\sc Yang, C., Wang, Q., Xu, M., Chen, Z., Bian, K., Liu, Y., and Liu, X.}
\newblock Characterizing impacts of heterogeneity in federated learning upon large-scale smartphone data.
\newblock In {\em {WWW} '21: The Web Conference 2021, Virtual Event / Ljubljana, Slovenia, April 19-23, 2021\/} (2021), J.~Leskovec, M.~Grobelnik, M.~Najork, J.~Tang, and L.~Zia, Eds., {ACM} / {IW3C2}, pp.~935--946.

\bibitem{YangLCT19Federated}
{\sc Yang, Q., Liu, Y., Chen, T., and Tong, Y.}
\newblock Federated machine learning: Concept and applications.
\newblock {\em {ACM} Trans. Intell. Syst. Technol. 10}, 2 (2019), 12:1--12:19.

\bibitem{ye2023heterogeneous}
{\sc Ye, M., Fang, X., Du, B., Yuen, P.~C., and Tao, D.}
\newblock Heterogeneous federated learning: State-of-the-art and research challenges.
\newblock {\em ACM Computing Surveys 56}, 3 (2023), 1--44.

\bibitem{YinCRB18Byzantine}
{\sc Yin, D., Chen, Y., Ramchandran, K., and Bartlett, P.~L.}
\newblock Byzantine-robust distributed learning: Towards optimal statistical rates.
\newblock In {\em Proceedings of the 35th International Conference on Machine Learning, {ICML} 2018, Stockholmsm{\"{a}}ssan, Stockholm, Sweden, July 10-15, 2018\/} (2018), J.~G. Dy and A.~Krause, Eds., vol.~80 of {\em Proceedings of Machine Learning Research}, {PMLR}, pp.~5636--5645.

\bibitem{fedlab}
{\sc Zeng, D., Liang, S., Hu, X., Wang, H., and Xu, Z.}
\newblock Fedlab: {A} flexible federated learning framework.
\newblock {\em J. Mach. Learn. Res. 24\/} (2023), 100:1--100:7.

\bibitem{Zeng23Tackling}
{\sc Zeng, D., Xu, Z., Pan, Y., Wang, Q., and Tang, X.}
\newblock Tackling hybrid heterogeneity on federated optimization via gradient diversity maximization.
\newblock {\em CoRR abs/2310.02702\/} (2023).

\bibitem{ZhangJCLW23A3FL}
{\sc Zhang, H., Jia, J., Chen, J., Lin, L., and Wu, D.}
\newblock {A3FL:} adversarially adaptive backdoor attacks to federated learning.
\newblock In {\em Advances in Neural Information Processing Systems 36: Annual Conference on Neural Information Processing Systems 2023, NeurIPS 2023, New Orleans, LA, USA, December 10 - 16, 2023\/} (2023), A.~Oh, T.~Naumann, A.~Globerson, K.~Saenko, M.~Hardt, and S.~Levine, Eds.

\bibitem{zhang2023pfllib}
{\sc Zhang, J., Liu, Y., Hua, Y., Wang, H., Song, T., Xue, Z., Ma, R., and Cao, J.}
\newblock Pfllib: Personalized federated learning algorithm library.
\newblock {\em arXiv preprint arXiv:2312.04992\/} (2023).

\bibitem{ZhangZL15}
{\sc Zhang, X., Zhao, J.~J., and LeCun, Y.}
\newblock Character-level convolutional networks for text classification.
\newblock In {\em Advances in Neural Information Processing Systems 28: Annual Conference on Neural Information Processing Systems 2015, December 7-12, 2015, Montreal, Quebec, Canada\/} (2015), C.~Cortes, N.~D. Lawrence, D.~D. Lee, M.~Sugiyama, and R.~Garnett, Eds., pp.~649--657.

\bibitem{ZhangPSYMMR022Neurotoxin}
{\sc Zhang, Z., Panda, A., Song, L., Yang, Y., Mahoney, M.~W., Mittal, P., Ramchandran, K., and Gonzalez, J.}
\newblock Neurotoxin: Durable backdoors in federated learning.
\newblock In {\em International Conference on Machine Learning, {ICML} 2022, 17-23 July 2022, Baltimore, Maryland, {USA}\/} (2022), K.~Chaudhuri, S.~Jegelka, L.~Song, C.~Szepesv{\'{a}}ri, G.~Niu, and S.~Sabato, Eds., vol.~162 of {\em Proceedings of Machine Learning Research}, {PMLR}, pp.~26429--26446.

\end{thebibliography}
}

\appendix

\begin{table}[tbh]
\tabcolsep=0.1cm
\renewcommand{\arraystretch}{1.2}
% \scriptsize
% \vspace*{-.5em}
\centering
\caption{Experimental setups of the FL benchmarks.}
\label{tab:exp_cfg}
\begin{tabular}{ccccc}
\toprule
\textbf{Domain} & \textbf{Dataset} & \textbf{Model} & \textbf{$T$} & \textbf{$\eta$} \\ 
\midrule
\midrule
\multirow{2}{*}{Image} & CIFAR10 & \multirow{2}{*}{ResNet18} & \multirow{2}{*}{2000} & 0.01 \\ 
% \cline{2-2} \cline{5-5}
 & FEMNIST &  &  & 0.001 \\ 
\cline{1-5}
\multirow{2}{*}{Text} & AGNews & \multirow{2}{*}{TinyBert} & \multirow{2}{*}{500} & \multirow{2}{*}{0.0001}  \\ 
% \cline{2-2}
 & IMDB &  &  &  \\ 
\cline{1-5}
\multirow{2}{*}{Tabular} & Purchase100 & \multirow{2}{*}{MLP} & \multirow{2}{*}{500} & \multirow{2}{*}{0.01}  \\ 
% \cline{2-2}
 & Texas100 &  &  &  \\ 
 \bottomrule
\end{tabular}
\end{table}

\section{Datasets and Models}
\label{sec:appendix_dataset_model}
\begin{itemize}
    \item CIFAR10~\cite{krizhevsky2009learning} dataset consists of 60,000 32×32 color images in 10 different classes, widely used for image classification tasks. Correspondingly, we use ResNet18~\cite{HeZRS16} as the backbone for image classification.
    \item FEMNIST~\cite{leaf} is a FL-oriented dataset based on the Extended MNIST (EMNIST) dataset containing handwritten characters. FEMNIST is designed to mimic the FL scenario, where the data is distributed across different clients or devices, and different clients have their own subsets of the FEMNIST data, which represents different handwritten characters and is partitioned in a way that reflects the heterogeneity of data across clients, similar to how real-world data from different users' handwritten notes might vary. For the backbone model, we use ResNet18 as well.
    \item Texas100 adopted in~\cite{ShokriSSS17Membership} includes hospital discharge data and information on inpatients from multiple medical institutions released by the Texas Department of State Health Services. The data records have details about external causes of injuries, diagnoses, and procedures that patients have undergone (such as surgeries), as well as general information like gender, age, race, hospital ID, and length of stay. The processed dataset contains 67,330 records and 6,170 binary features, representing the 100 most common medical procedures. These records are divided into 100 categories, each representing different types of patients for classification models. For this tabular data, we normalize the data and use MLP for classification (the same as Purchase100).
    \item Purchase100 adopted in~\cite{ShokriSSS17Membership} is the shopping history data of thousands of people. Each user's record encompasses their transaction records within a year, including product names, stores, quantities, and dates, with 197,324 pieces in the simplified version. Each piece of data consists of 600 binary digits, with each digit indicating whether a certain product has been purchased or not. 
    \item AGNews~\cite{ZhangZL15} is a well-known text-classification dataset. It consists of news articles from the AG (Associated Press) news agency. The dataset evaluates algorithms for text classification tasks, such as categorizing news articles into different topics. It contains four different classes of news: World, Sports, Business, and Sci/Tech. This class distribution allows for a multi-class classification problem setup. For this task, we use pre-trained TinyBert~\cite{JiaoYSJCL0L20} as the backbone (the same as IMDB).
    \item IMDB~\cite{imdb} (Internet Movie Database) dataset is a well-known resource for movie-related data, containing movie reviews and ratings with two categories for sentiment analysis. The reviews are usually in text format and are accompanied by a numerical rating given by users. 
\end{itemize}

\section{Evaluated Attacks}
\label{sec:appendix_evaluated_attacks}
\subsection{Backdoor Attack}
\begin{itemize}
    \item Model replace backdoor attack (MRBA)~\cite{BagdasaryanVHES20How} aims to ensure that the adversary's malicious updates (backdoor task) survive the aggregation process used in FL, such as FedAvg. The attacker achieves this by scaling up the magnitude of their local model update to dominate the global model update. This approach allows the adversary to effectively inject a backdoor into the global model. The success of this method lies in its ability to bypass defenses that might otherwise mitigate the impact of malicious updates by diluting them during aggregation.
    \item Distributed backdoor attack (DBA)~\cite{xie2019dba} presents a novel form of backdoor attack. Unlike centralized backdoor attacks, where all malicious parties use the same global trigger pattern, DBA decomposes the global trigger into separate local triggers embedded into the training data of distinct adversarial clients. This distributed nature makes DBA more persistent and stealthy than centralized attacks. Extensive experiments demonstrate that DBA significantly outperforms centralized backdoor attacks in terms of attack success rate while evading robust FL algorithms designed to defend against centralized backdoors.
    \item Edgecase backdoor attack (Edgecase)~\cite{WangSRVASLP20Attack} introduces an edge-case backdoor, which targets seemingly simple inputs unlikely to appear in the training or test data. These inputs reside at the tail end of the data distribution, making them less likely to be encountered naturally. By carefully tuning the attack parameters, adversaries can insert these edge-case backdoors across multiple tasks without noticeably affecting the main task performance. This subtle manipulation can lead to undesirable failures and serious repercussions on fairness.
    \item Neurtoxin backdoor attack (Neurtoxin)~\cite{ZhangPSYMMR022Neurotoxin} focuses on creating a durable backdoor that persists even after many rounds of FL training by exploiting the fact that some parts of neural networks may not change much over time, allowing the backdoor to remain effective. Neurotoxin uses a combination of techniques, including gradient alignment and selective poisoning, to ensure that the backdoor remains active despite changes in the global model.
    \item Cerp backdoor attack (CerP)~\cite{LyuHWLWL023Poisoning} describes an attack that leverages collusion between multiple attackers to create a more potent and harder-to-detect backdoor. CerP involves jointly crafting poison samples that, when aggregated, result in a powerful backdoor effect. The stealthiness comes from the fact that individual contributions from each colluding party appear benign, but collectively, they form a damaging influence on the global model.
    \item A3FL backdoor attack (A3FL)~\cite{ZhangJCLW23A3FL} proposes an adversarially adaptive backdoor attack framework. A3FL dynamically adjusts its strategy based on the defense mechanisms deployed within the FL system. It does so by continuously monitoring the effectiveness of its attacks and adapting accordingly, ensuring that the backdoor remains effective even in the presence of evolving defenses.
\end{itemize}

\subsection{Byzantine Attack}

\begin{itemize}
    \item Fang attack~\cite{FangCJG20Local} is an aggregator-specific attack involving collusive behavior, where each adversary is aware of the other adversaries’ local training data, models, and gradient updates. If the aggregator is unknown, a randomly assumed aggregator is used. Here we present the Krum-version Fang attack, which deviates the update while optimizing it to be selected by Krum as if it were benign.
    \item IPM attack~\cite{XieKG19Fall} observes that many robust aggregators only control the distance between the aggregated update and the benign mean, but do not guarantee that the aggregated direction remains aligned with the true descent direction. The attack therefore crafts malicious updates that reverse the inner product between the robust aggregate and the benign gradient, pushing training away from convergence.
    \item LabelFlip attack~\cite{KarimireddyHJ21Learning} alters training labels according to an attacker-chosen substitution rule, which can produce random, inverse, or target labels. In the random strategy, the source label is replaced with a random one; in the inverse strategy, label $l$ is flipped to $L-l-1$, where $L$ is the number of classes; and in the target strategy, a specific label is assigned to label $l$. It supports both targeted and untargeted poisoning, and the targeted version is used in our paper.
    \item SignFlip attack~\cite{LiXCGL19RSA} aims to deviate the global model in the wrong direction by flipping the signs of local gradient updates. Compared with random perturbation, this attack preserves the magnitude scale of the update while consistently steering optimization toward ascent rather than descent.
    \item UpdateFlip attack operates at the update level instead of only the coordinate-sign level: the adversary reverses the direction of the local model update before submission so that aggregation moves the global model away from the benign trajectory. In our benchmark, it serves as a simple but effective directional corruption baseline.
    \item MinMax attack~\cite{ShejwalkarH21Manipulating} searches for a malicious update whose maximum distance to benign updates stays below the largest pairwise benign distance. By remaining within the benign spread while still shifting the aggregate, the attack is able to degrade learning without appearing as an obvious outlier.
    \item LIE attack~\cite{BaruchBG19A} identifies that instead of relying on large parameter changes, carefully crafted small perturbations from only a few clients can already undermine convergence and bypass defenses such as Krum, TrimmedMean, and Bulyan. It estimates the benign mean and variance coordinate-wise and places adversarial updates in a statistically plausible region so that the attack remains hard to remove while still biasing aggregation.
    \item MedianTailored attack is a coordinate-wise attack tailored to median-based robust aggregation. Rather than maximizing the norm of the perturbation, it crafts each coordinate to shift the coordinate-wise median toward an adversarial value while staying inside a range that still appears compatible with benign updates.
    \item Noise attack~\cite{LiNV23Byzantine} injects random perturbations into local gradient updates to disrupt training, particularly in later rounds when benign updates become more concentrated. In our implementation, each adversary submits Gaussian noise vectors whose mean and variance control the strength of the attack.
    \item SignGuard attack~\cite{XuHSL22Byzantine} is included in our benchmark as a sign-aware malicious-update setting targeting collaborative gradient filtering. In this setting, adversarial clients manipulate the sign pattern and magnitude structure of their submitted updates so as to reduce utility while remaining less conspicuous to sign-based filtering mechanisms.
\end{itemize}

\section{Experimental Configuration}

\subsection{FL Benchmark Configuration}
Tab.~\ref{tab:exp_cfg} summarizes the benchmark-specific training horizon and learning rate used in our experiments. In all benchmarks, unless otherwise stated, we use $N=100$ total clients, join ratio $\alpha=0.1$, local batch size $B=64$, and local epoch number $E=5$. We hold out 10\% of each dataset for FL global evaluation, while the remaining data are distributed across clients for federated training. FedAvg is used as the default aggregation algorithm throughout the main experiments.

For the benchmark choices, we evaluate image, text, and tabular FL tasks with dataset-model pairs that match the corresponding application modality: CIFAR10 and FEMNIST with ResNet18, AGNews and IMDB with TinyBert, and Purchase100 and Texas100 with MLP. The training horizon $T$ and learning rate $\eta$ follow Tab.~\ref{tab:exp_cfg}. For text benchmarks, we use $T=500$ and $\eta=10^{-4}$; for tabular benchmarks, we use $T=500$ and $\eta=10^{-2}$; for image benchmarks, we use $T=2000$ with $\eta=10^{-2}$ for CIFAR10 and $\eta=10^{-3}$ for FEMNIST.

We evaluate two FL conditions. The \emph{practical} setting uses hybrid heterogeneity with $\tau^{s}=0.9$, $\tau^{c}=0.9$, and $\tau^{d}=0.9$, while the \emph{ideal} setting uses $\tau^{s}=\infty$, $\tau^{c}=1$, and $\tau^{d}=1$ (except for FEMNIST, where the original client partition is retained). In the security-only arXiv version, we do not enable additional poisoning-robust defenses in the main benchmark so that the reported results isolate the impact of threat-model and evaluation gaps.

\subsection{Attack Configuration}
Our poisoning benchmark includes both backdoor and byzantine attacks. For backdoor evaluation, we implement DBA, CerP, EdgeCase, A3FL, Model Replacement, and Neurotoxin when the attack is compatible with the target modality. In the main backdoor figures, the poison ratio is swept over $\{1,3,5,7,10\}\%$. For image benchmarks, we report the full image-compatible set of attacks; for text and tabular benchmarks, we report the subset that is directly implementable in our framework, namely DBA, EdgeCase, Replace, and Neurotoxin.

For byzantine evaluation, we include IPM, Noise, Fang, LabelFlip, SignFlip, UpdateFlip, MinMax, LIE, MedianTailored, and SignGuard. Across attacks, we align the threat model with the practical assumptions in Sec.~\ref{sec:gap1}: malicious clients are sampled through the same random participation process as benign clients, the attacker does not assume fixed aggregation participation, and malicious clients do not rely on direct access to benign updates when such access is unrealistic for deployment.

Attack-specific internal hyperparameters are intentionally not exhaustively enumerated in this arXiv version. In particular, the exact trigger pattern, target label, local poison fraction, scaling coefficients, and optimization-side attack coefficients are left unspecified here when they differ across attacks or were adjusted during implementation. The goal of this appendix is therefore to document the common evaluation protocol and the attack families covered by \textit{\textbf{TFLlib}}, while leaving method-specific tuning details to a subsequent full release.

\subsection{Detailed Performance}
\begin{figure*}
\centering
\includegraphics[width=0.99\linewidth]{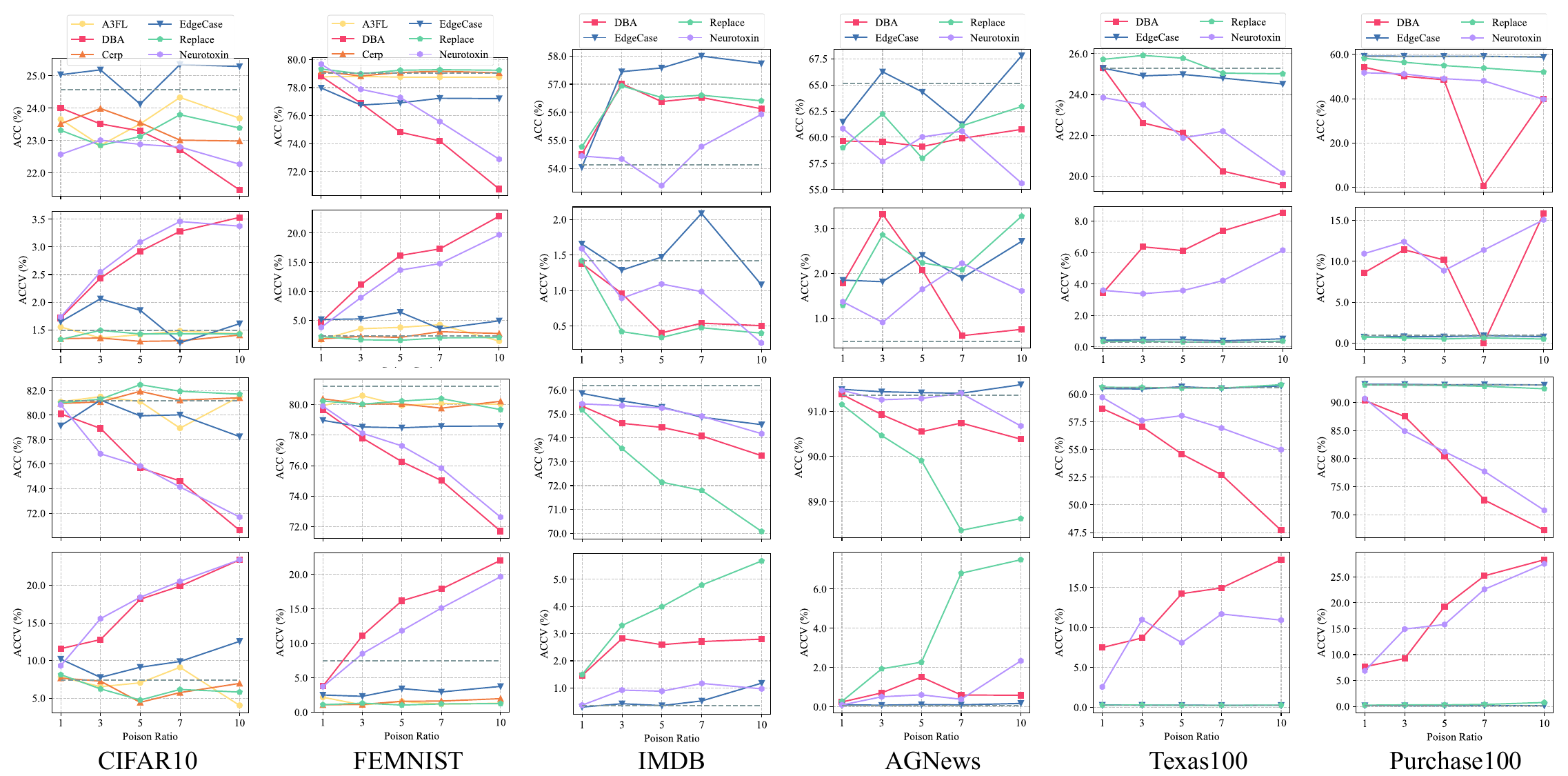}
\caption{Backdoor Stealthiness under Practical (\textit{1st line}) and Ideal (\textit{2nd line}) Settings.}
\label{fig:main_bkd_acc}
\end{figure*}

The detailed backdoor stealthiness results corresponding to the main-text discussion are shown in Fig.~\ref{fig:main_bkd_acc}. The top two rows report the practical setting and the bottom two rows report the ideal setting. For each dataset, the figure complements Fig.~\ref{fig:main_bkd_asr} in the main text by showing the benign-task accuracy and its variance under different poison ratios, which allows us to judge whether a seemingly effective attack remains operationally stealthy.

\end{document}